\documentclass[prd,aps,twocolumn,a4paper,floatfix,nofootinbib]{revtex4}

\usepackage{graphicx,psfrag}
\usepackage{mathrsfs}
\usepackage{amsmath,amsfonts,amssymb}
\usepackage{multirow}
\usepackage{hyperref}
\usepackage{placeins}
\usepackage[normalem]{ulem}
\usepackage{slashed} 
\usepackage{amsthm}
\usepackage{latexsym}
\usepackage[dvips]{epsfig}

\begin{document}
%%%%%%%%%%%%%%%%%%%%%%%%%%%%%%%%%%%%%%%%%%%%%%%%%%%%%%%%%%%%%%%%%%%%%%%%%%%%%%%%%%%%%%%

\title{The Weak Null Condition in Free-evolution Schemes for Numerical Relativity:
  \newline Dual Foliation GHG with Constraint Damping}

\author{Edgar \surname{Gasper\'in} and David \surname{Hilditch}}

\affiliation{Centro de Astrof\'isica e Gravita\c c\~ao - CENTRA,
  Departamento de F\'isica, Instituto Superior T\'ecnico - IST,
  Universidade de Lisboa - UL, Av. Rovisco Pais 1, 1049-001 Lisboa,
  Portugal}

\date{\today}

\begin{abstract}
All strategies for the treatment of future null-infinity in numerical
relativity involve some form of regularization of the field
equations. In a recent proposal that relies on the dual foliation
formalism this is achieved by the use of an asymptotically Minkowskian
generalized harmonic tensor basis. For the scheme to work however,
derivatives of certain coordinate light-speeds must decay fast
enough. Presently, we generalize the method of asymptotic expansions
for nonlinear wave equations to treat first order symmetric hyperbolic
systems. We then use this heuristic tool to extract the expected rates
of decay of the metric near null-infinity in a free-evolution
setting. We show, within the asymptotic expansion, that by carefully
modifying the non-principal part of the field equations by the
addition of constraints, we are able to obtain optimal decay rates
even when the constraints are violated. The light-speed condition can
hence be satisfied, which paves the way for the explicit numerical
treatment of future null-infinity. We then study the behavior of the
Trautman-Bondi mass under the decay results predicted by the
asymptotic expansion. Naively the mass seems to be unbounded, but we
see first that the divergent terms can be replaced with a combination
of the constraints and the Einstein field equations, and second that
the Bondi mass loss formula is recovered within the framework. Both of
the latter results hold in the presence of small constraint
violations.
\end{abstract}

\maketitle

\tableofcontents

%%%%%%%%%%%%%%%%%%%%%%%%%%%%%%%%%%%%%%%%%%%%%%%%%%%%%%%%%%%%%%%%%%%%%%%%%%%%%%%%%%%%%%%
\section{Introduction}\label{Section:Introduction}
%%%%%%%%%%%%%%%%%%%%%%%%%%%%%%%%%%%%%%%%%%%%%%%%%%%%%%%%%%%%%%%%%%%%%%%%%%%%%%%%%%%%%%%

The notion of gravitational radiation is intimately related to the
asymptotic behavior of the gravitational field. According to Penrose's
proposal~\cite{Pen63, Pen64, Hus02} isolated systems may be described
by asymptotically simple spacetimes. Among these spacetimes those with
vanishing cosmological constant, also known as \emph{Minkowski-like
  spacetimes}, play a fundamental role for the current understanding
of gravitational radiation~\cite{Ste91, Fra04, Ash14, Val16}. For
these spacetimes, the conformal boundary~$\mathscr{I}$ is a
null-hypersurface representing idealized observers at infinity. It is
in this setting, in which the concept of gravitational radiation,
originally introduced at the linear level, can be made rigorous in the
non-linear theory and results such as the loss of mass-energy due to
outgoing gravitational radiation~\cite{Tra58a, BonBurMet62, Sac62a}
properly formulated.

Despite the advances achieved in numerical relativity (NR) in the last
two decades, the inclusion of future null infinity in the
computational domain is in general still an open problem. In view of
the first direct detection of gravitational waves in
2016~\cite{AbbAbbAbb16} and the further development of gravitational
wave astronomy in the years to come, the most natural motivation to
solve this problem is for the computation of astrophysical wave
forms at future null infinity. Nevertheless there are other important
principle reasons to include the conformal boundary in the numerical
domain, not least the weak cosmic censorship conjecture.

There are several approaches to include the conformal boundary in the
computational domain. Possible avenues include Cauchy-Characteristic
Matching~\cite{Win12} and the use of a suitable hyperboloidal initial
value problem. In the latter approach, which we follow, initial data
is given on slices which are everywhere spacelike but which terminate
at future null infinity. This is combined with a suitable radial
compactification. Naively looking at the field equations on such a
slice however we get singular expressions, for which some cure is
needed, particularly for numerical applications. One option is to use
the conformal Einstein field equations introduced by Friedrich
in~\cite{Fri81}. Although this approach is well suited to study the
asymptotic region of the spacetime, it poses difficulties to
numerically evolve realistic astrophysical scenarios, as it is not
clear how to adapt those codes and techniques that already work well
in the strong field region, for instance for compact binaries. Recent
progress on the numerical implementation of the conformal Einstein
field equations has however been reported in~\cite{DouFra16}. Another
proposal is to use a standard formulation of GR in generalized
harmonic gauge (GHG) but insist on performing a full compactification
of spacetime. This leads to equations which are formally singular but
which take a finite limit in a suitable gauge~\cite{Zen08}. Following
up on these ideas, Va\~n\'o-Vi\~nuales and collaborators have shown
successful numerical evolutions in spherical
symmetry~\cite{VanHusHil14,VanHus14,Van15}. See also~\cite{MonRin08,
  BarSarBuc11} for other related approaches.

An alternative proposal, using the coordinates of~\cite{CalGunHil05},
was given in~\cite{HilHarBug16}. Here the strategy, broadly speaking,
is to pose an hyperboloidal initial value problem and exploit the dual
foliation formalism~\cite{Hil15}. In the dual foliation formalism two
coordinate systems with two different foliations are employed. This
allows us to use one coordinate system to construct a global tensor
basis and another to coordinatize the spacetime. In the current
application, it is natural to take the first coordinate system adapted
to a Cauchy hypersurface, and the other to a hyperboloidal
slice~\cite{HilHarBug16}. In this proposal, one employs a first order
reduction of GHG. The regularization strategy is based on the idea
that, near null infinity, the metric looks trivial in the GHG tensor
basis. It was found in~\cite{HilHarBug16} that for the regularization
to work certain derivatives of the coordinate light-speeds must have
enough decay. We call this the \emph{coordinate light-speed
  condition}.

In this article we analyze whether or not this condition can be
realized and if so, in which circumstances. Interestingly, this is
connected with the the weak null condition~\cite{LinRod03} and,
consequently, to the concept of \emph{asymptotic systems}. The
asymptotic system can be regarded as a heuristic tool to predict the
asymptotics of the solutions to a general a system of quasilinear wave
equations. We view this tool as heuristic because, to the best of our
knowledge, it has not yet been established that solutions to the
asymptotic system will always have the same asymptotics as solutions
to the original system.

It has however been shown that if a system of wave equations satisfy
the \emph{hierarchical weak null condition}, a condition slightly
stronger than the weak null condition, then, the original PDEs admit
global solutions whose asymptotics agree with the prediction of the
asymptotic system~\cite{Kei17}. An important consequence is that
global existence for GR in harmonic gauge can be established without
assuming that the constraints are satisfied. Moreover when the
constraints are satisfied the formulation of~\cite{LinRod03} is
expected to satisfy the coordinate light-speed condition, so naively
the situation looks promising. In free evolution schemes however,
constraint violations are not only present but expected to grow
rapidly, rendering the solution
unphysical~\cite{BroFriHub98,GunGarCal05,LinSchKid05}. Therefore the
use of the field equations as in~\cite{LinRod03} is not appropriate
for numerical free evolution as envisioned in our program. This is an
old problem in NR, and various approaches have been suggested to
alleviate it. For instance in~\cite{GunGarCal05} a constraint addition
was proposed which effectively damps away high frequency violations at
the linear level. Unfortunately for our desired application the
formulation of~\cite{LinRod03,Kei17} is {\it not} expected to satisfy
the coordinate light-speed condition in the presence constraint
violations. Presently we solve the problem with a GHG formulation,
making a careful constraint addition to the field equations. With this
addition the corresponding asymptotic system predicts that constraint
violations are strongly damped as one approaches null infinity, and we
recover the same fall-off present in their absence. This in turn
should ensure that the coordinate light-speed condition
of~\cite{HilHarBug16} is satisfied even in the presence of small
constraint violations.

Although the calculation of the asymptotic system is straightforward,
most of the literature about asymptotic
systems~\cite{Hor87,Hor97,LinRod03} has been given in the context of
second order equations. In view of the fact that the formulation of
the Einstein field equations used in~\cite{HilHarBug16} is first
order, we give a discussion of how to compute the asymptotic system
for such systems. Finally we discuss the connection between the
Trautman-Bondi mass and the asymptotic system and recover the mass
loss formula in this context. We also see that modifying the
definition of the Trautman-Bondi mass with the constraints, similar
results may be obtained even when small constraint violations are
present.

%%%%%%%%%%%%%%%%%%%%%%%%%%%%%%%%%%%%%%%%%%%%%%%%%%%%%%%%%%%%%%%%%%%%%%%%%%%%%%%%%%%%%%%
\section{Basic set up}\label{BasicSetUp}
%%%%%%%%%%%%%%%%%%%%%%%%%%%%%%%%%%%%%%%%%%%%%%%%%%%%%%%%%%%%%%%%%%%%%%%%%%%%%%%%%%%%%%%

In what follows, lower case Latin letters from the first half of the
alphabet will denote abstract indices while Greek letters will be used
to denote coordinate indices. In addition, capital Latin letters will
be use to label the elements of a non-coordinate frame basis (the
angular part of a spacetime null frame).  Given a 2-tensor~$T_{ab}$
and arbitrary frame vector fields~$X^a$, $Y^b$ the components
of~$T_{ab}$ in this frame will be denoted as~$T_{XX}=X^aX^bT_{ab}$,
$T_{XY}=X^aY^bT_{ab}$, $T_{YX}=Y^aX^bT_{ab}$ and~$T_{YY}=Y^aY^bT_{ab}$
and similarly for higher-valence tensors. Let~$m_{ab}$
and~$\mathring{\nabla}$ represent the Minkowski metric and its
corresponding Levi-Civita connection.  Let~$(T,R,\theta^{A})$ denote
spherical polar coordinates where~$\theta^{A}$ with~$A \in \{ 1,2\}$
represent some arbitrary coordinates on~$\mathbb{S}^2$. Let~$\omega_{
  A}{}^{a}$ with~$ A \in \{ 1,2 \}$ denote a frame with corresponding
coframe~$\hat{\omega}^{ A}{}_{a}$, such that
\begin{align}
  \sigma_{ab}=\delta_{ A B}\hat{\omega}^{ A}{}_{a}\hat{\omega}^{ B}{}_{b},
\end{align}
where~$\sigma_{ab}$ represents the standard metric on~$\mathbb{S}^2$
and~$\delta_{ A B}$ is the Kronecker delta. Observe
that,~$\omega_{ A}{}^{a}$ is not a coordinate frame. In fact,
respect to the natural coordinate basis one can write
\begin{align}
\hat{\omega}^{ A}{}_{a}=\Omega_{B}{}^{ A}(\mbox{d}\theta^B)_a,
\end{align}
where~$\Omega_{B}{}^{ A}$ is a matrix which only depends on the
coordinates~$\theta^A$. In this frame, the Minkowski metric can be
expressed in abstract index notation as
\begin{align}  
m_{ab}= -(\mbox{d}T)_a(\mbox{d}T)_{b}+(\mbox{d}R)_a(\mbox{d}R)_{b}+
R^2\sigma_{ab}.
\end{align}
Let~$(L^{a},\underline{L}^a, S_{ A}{}^a)$ denote a null frame
respect to~$m_{ab}$ with the following normalization
\begin{align}
 m_{L\underline{L}} = -2, \qquad \nonumber 
 m_{S_{ A}S_{ B}}=\delta_{ A B}, \quad \nonumber  \\
 m_{LL}=m_{\underline{L}\underline{L}}=m_{LS_{ A}}=m_{\underline{L}S_{ A}}=0.
 \end{align}
The relation between the \emph{flat null
 frame}~$(L^{a},\underline{L}^a, S_{ A}{}^a)$
and~$((\partial_T)^a,(\partial_R)^a,\omega_{ A}{}^a)$ is given by
\begin{flalign}\label{eq:NullFrameToSpFrame}
L^a = & \; (\partial_T)^a + (\partial_R)^a, \qquad
\underline{L}^a=(\partial_T)^a - (\partial_R)^a, \nonumber\\
S_{ A}{}^a= & \; \frac{1}{R}\omega_{ A}{}^{a}.
\end{flalign}
Similarly, let~$(\hat{L}_{a},\underline{\hat{L}}_a, \hat{S}^{ A}{}_a)$
denote the associated coframe. Observe that in terms of the corresponding
coframe~$((\mbox{d}T)_a,(\mbox{d}R)_a,\omega^{ A}{}_a)$ one has
\begin{flalign}\label{eq:NullcoFrameToSpcoFrame}
 \hat{L}_a = & -(\mbox{d}T)_a + (\mbox{d}R)_a, \qquad
 \underline{\hat{L}}_a= -(\mbox{d}T)_a - (\mbox{d}R)_a,
 \nonumber\\ \hat{S}^{ A}{}_{a}= & \; R\hat{\omega}^{ A}{}_a.
\end{flalign}
Therefore, the Minkowski metric can be succinctly expressed as
\begin{align}
  m_{ab}=-\hat{L}_{(a}\underline{\hat{L}}_{b)}+\delta_{ A B}
  \hat{S}^{ A}{}_{a}\hat{S}^{ B}{}_{b}.
\end{align}

More generally, observe that, given a symmetric
tensor~$T_{ab}=T_{(ab)}$ one can express it as,
\begin{flalign}\label{eq:SymmetricTensorInFlatNullFrame}
 T_{ab} & = \tfrac{1}{4} T_{\underline{L} \underline{L}}{} \hat{L}_{a} \hat{L}_{b}
+ \tfrac{1}{2} T_{L\underline{L}}{} \hat{L}_{(a} \underline{\hat{L}}_{b)} +
\tfrac{1}{4} T_{LL}{} \underline{\hat{L}}_{a} \underline{\hat{L}}_{b}
+ T^{\textrm{TF}}_{AB}\perp^{AB}_{ab} \nonumber\\&
+ \tfrac{1}{2}T^{\varnothing}\delta_{ A B}\hat{S}^{ A}{}_{a}\hat{S}^{ B}{}_{b} 
- T_{\underline{L} S_{ A}}{} \hat{L}_{(a} \hat{S}^{ A}{}_{b)} -
T_{LS_{ A}}{} \underline{\hat{L}}_{(a} \hat{S}^{ A}{}_{b)},
\end{flalign}
where we have defined the trace on the sphere,
\begin{align}\label{TraceonSphere}
T^{\varnothing}\equiv \delta^{ A B}T_{S_{ A}S_{ B}},
\end{align}
and the tracefree part of the projection,
\begin{align}
T^{\textrm{TF}}_{AB} &=\perp^{ab}_{AB}T_{ab}.
\end{align}
For this we employ the tracefree projection operator,
\begin{align}
  \perp^{ab}_{AB}&=\hat{S}_A{}^a\hat{S}_B{}^b
  -\tfrac{1}{2}\delta_{AB}(\delta^{CD}\hat{S}_C{}^a\hat{S}_D{}^b), \nonumber \\
  \perp^{AB}_{ab}&=\hat{S}^A{}_a\hat{S}^B{}_b
  -\tfrac{1}{2}\delta^{AB}(\delta_{CD}\hat{S}^C{}_a\hat{S}^D{}_b).
\end{align}
The independent components of~$T^{\textrm{TF}}_{AB}$ are
\begin{align}
T^{\textrm{TF}}_{11}&=-T^{\textrm{TF}}_{22}=\frac{1}{2}(T_{S_1S_1}-T_{S_2S_2})\equiv
T^{+}, \nonumber \\ T^{\textrm{TF}}_{12}&=T_{S_1S_2}\equiv T^{\times}
\label{TracefreePlusCross}.
\end{align}
They correspond of course to the two gravitational wave polarizations
in the case of physical interest.

%%%%%%%%%%%%%%%%%%%%%%%%%%%%%%%%%%%%%%%%%%%%%%%%%%%%%%%%%%%%%%%%%%%%%%%%%%%%%%%%%%%%%%%
\section{The asymptotic system for first and second order systems}
%%%%%%%%%%%%%%%%%%%%%%%%%%%%%%%%%%%%%%%%%%%%%%%%%%%%%%%%%%%%%%%%%%%%%%%%%%%%%%%%%%%%%%%

%%%%%%%%%%%%%%%%%%%%%%%%%%%%%%%%%%%%%%%%%%%%%%%%%%%%%%%%%%%%%%%%%%%%%%%%%%%%%%%%%%%%%%%
\subsection{Second order systems} \label{SecondOrderAsymptotic}
%%%%%%%%%%%%%%%%%%%%%%%%%%%%%%%%%%%%%%%%%%%%%%%%%%%%%%%%%%%%%%%%%%%%%%%%%%%%%%%%%%%%%%%

Consider a set of quasilinear wave equations of the form
\begin{flalign}
\mathring{\square} u_{I} = \sum_{
  \substack{| \alpha|\leq| \beta|\leq2,\\ | \beta| \geq
    1}}a^{JK}_{I, \alpha \beta}\partial^{ \alpha}
u_{J}\partial^{ \beta} u_{K} + G(u,u',u''),\label{eq:WaveEqForDefWeakNull}
\end{flalign}
where~$I,J,K \in \{1,...,n\}$, $u=(u_1,...,u_n)$, $\mathring{\square}=
-\partial_t^2+\delta^{ij}\partial_{i}\partial_j$, $x^i$ with $i,j \in
\{1,2,3\}$, $ \alpha$, $ \beta$ are multi-indices and
$G(u,u',u'')$ vanishes to third order as $(u,u',u'')\rightarrow 0$,
with small initial data
\begin{align}
u(0,x)=v(x)\in C^\infty, \quad \partial_Tu(0,x)=w(x)\in C^\infty,
\end{align}
decaying fast as~$R\rightarrow\infty$, where~$R=
\sqrt{\delta_{ij}x^ix^j}$. Recall that a quadratic
form~$N_{ab}\equiv~N(\mathring{\nabla}_a w,\mathring{\nabla}_a w)$
where~$ w$ is a tensor, is said to be a null form if it vanishes upon
formal replacement of~$\mathring{\nabla}_a w$ with $v_a w$,
where~$v_a$ is null. Similar definitions are made for quadratic forms
involving higher derivatives. The set~\eqref{eq:WaveEqForDefWeakNull}
is said to satisfy the null-condition if the quadratic nonlinearity is
a null-form. The value of this definition is that,
in~$3+1$-dimensions, systems that satisfy the null-condition have
global solutions when fed small data, and moreover have solutions of
the same asymptotic behavior as a linear wave equation near
null-infinity~\cite{Sog95}. The key subtlety is that the
null-condition is sufficient for long-term existence, but not
necessary. For numerical applications the more relevant question
concerns the asymptotic behavior of solutions rather than small-data
global existence, but because of their close relationship in the
mathematics literature they are often discussed in tandem. For example
the weak null condition, originally introduced in~\cite{LinRod03}, is
crucially tied to the notion of the \emph{asymptotic system}.  One
says that the wave equation~\eqref{eq:WaveEqForDefWeakNull} satisfies
the weak null condition if the corresponding asymptotic system,
\begin{flalign}
& \partial_s\partial_q U_I = A^{JK}_{I,nm}(\varpi)\partial^m_q U_J
\partial^n_q U_K,
\end{flalign}
where 
\begin{flalign}
& A^{JK}_{I,nm}(\varpi)=\sum_{| \alpha|=n, | \beta|=m}
a_{I, \alpha \beta}\hat{\varpi}^{ \alpha}\hat{\varpi}^{ \beta}, \nonumber\\
& \quad \quad \hat{\varpi}=(-1,\varpi), \qquad \varpi \in \mathbb{S}^2,
\end{flalign}
and~$U_i= R u_i$,~$q=R-T$,~$s=\ln R$, has solutions which, roughly
speaking, exist for all~$s$, and whose derivatives grow at most
exponentially in~$s$ for all initial data decaying sufficiently fast
in~$q$. In this language, the classical null condition states
that~$A^{JK}_{I,nm}(\varpi)=0$. Thus, a system of quasilinear wave
equations of the form given above satisfying the classical null
condition will trivially satisfy the weak null condition,
see~\cite{LinRod03, LinRod05, Lin17} for further details.  It is
conjectured that systems satisfying the weak null condition admit
small-data global solutions with asymptotics near null-infinity as
predicted by the asymptotic system.

%%%%%%%%%%%%%%%%%%%%%%%%%%%%%%%%%%%%%%%%%%%%%%%%%%%%%%%%%%%%%%%%%%%%%%%%%%%%%%%%%%%%%%%
\paragraph*{Model equation in second order form:}
%%%%%%%%%%%%%%%%%%%%%%%%%%%%%%%%%%%%%%%%%%%%%%%%%%%%%%%%%%%%%%%%%%%%%%%%%%%%%%%%%%%%%%%

An illustrative example is to consider the following~\emph{model
  equation}
\begin{flalign}\label{ToyModelIII}
\mathring{\square}\phi=(\partial_T \tilde{\phi})^2, \qquad
\mathring{\square}\tilde{\phi}=0.
 \end{flalign}
where the flat wave operator~$\mathring{\square}$ is expressed in
spherical polar coordinates~$(T,R,\theta^A)$.  To derive the
asymptotic system, define~$\Phi =R \phi$ and~$\tilde{\Phi}= R
\tilde{\phi}$ and make the change of coordinates~$(T,R,\theta^A)
\rightarrow (q,s,\theta^A)$ where~$q=R-T$ and~$s= \ln R$. Substituting
the expression for~$\phi$ and~$\tilde{\phi}$ in terms of the rescaled
variables~$\Phi$ and~$\tilde{\Phi}$ into equation~\eqref{ToyModelIII}
and formally equating the terms with coefficients~$R^{-2}$, one obtains
the following asymptotic system
\begin{flalign}\label{AsymptoticToySecondOrder}
\partial_s\partial_q \Phi= (\partial_q \tilde{\Phi})^2, \qquad
\partial_s\partial_q \tilde{\Phi}=0.
\end{flalign}
Equation~\eqref{ToyModelIII} is an example of a system of equations
that does not satisfy the classical null condition but satisfies the
weak null condition.  Observe that the second equation
in~\eqref{AsymptoticToySecondOrder} simply states that~$\partial_
q\tilde{\Phi}$ does not depend on~$s$. Then, using the method of
characteristics, one can integrate in~$s$ the first equation
in~\eqref{AsymptoticToySecondOrder} to conclude that~$\partial_q\Phi=
s (\partial_q\tilde{\Phi})^2$. The latter implies for the original set
of variables that~$\partial \phi$ decays as~$R^{-1}\ln
R$~\cite{LinRod03}.

The motivation for the introduction of the asymptotic system is
related to the observation that one expects better decay rates
for~$L$-derivatives of the field compared with other derivatives.
Therefore, the construction of the asymptotic system is essentially to
rescale the raw variables according to their expected fall-off, move
to suitable coordinates, and then to throw away all quadratic terms
containing derivatives containing at least one derivative tangential
to the outgoing light-cone, along with cubic and higher order terms.
We will see in the next section how each of these ingredients is
adjusted for first order systems.

%%%%%%%%%%%%%%%%%%%%%%%%%%%%%%%%%%%%%%%%%%%%%%%%%%%%%%%%%%%%%%%%%%%%%%%%%%%%%%%%%%%%%%%
\subsection{First order systems}\label{sec:FirstOrderSysAsympt}
%%%%%%%%%%%%%%%%%%%%%%%%%%%%%%%%%%%%%%%%%%%%%%%%%%%%%%%%%%%%%%%%%%%%%%%%%%%%%%%%%%%%%%%

To the best of our knowledge, the notion of the weak null condition
and the asymptotic systems has been exploited only for systems of wave
equations. In the context of gravitation, in particular, for the
Einstein field equations in harmonic gauge in~\cite{LinRod04} and more
general systems of wave equations in~\cite{Kei17}. Although the wave
equation can be recast straightforwardly as a first order symmetric
hyperbolic system by the introduction of an order reduction
variable~$w=\partial u$, the weak null condition and the notion of the
asymptotic system has not been discussed in this context yet.  This is
of importance for applications since numerical evolution schemes
often make use of a first order formulation of the field
equations. Moreover, although mathematically, the relation~$w=\partial
u$ connects the first and second order formulations, numerically, the
use of reduction variable requires the introduction of a reduction
constraint~$\mathcal{C} \equiv w-\partial u$, and violations to this
reduction constraint tends to grow exponentially, or worse, during
numerical
evolutions~\cite{BroFriHub98,GunGarCal05,LinSchKid05,WeyBerHil11}. Therefore,
one is forced to modify the evolution equations to control this bad
behavior.  With this motivation in mind, in this section, the model
equation~\eqref{ToyModelIII} is written in a form that is suitable for
numerical implementations and then the corresponding asymptotic system
is derived.

%%%%%%%%%%%%%%%%%%%%%%%%%%%%%%%%%%%%%%%%%%%%%%%%%%%%%%%%%%%%%%%%%%%%%%%%%%%%%%%%%%%%%%%
\paragraph*{Model equation in first order form:}
%%%%%%%%%%%%%%%%%%%%%%%%%%%%%%%%%%%%%%%%%%%%%%%%%%%%%%%%%%%%%%%%%%%%%%%%%%%%%%%%%%%%%%%

Let~$(\mathring{\mathcal{M}},m_{ab})$ denote the Minkowski spacetime
equipped with the objects introduced in section~\ref{BasicSetUp} and
consider the model equation~\eqref{ToyModelIII}. Let~$\mathring{N}^a$
denote the time unit normal to the surface determined by the
condition~$T=0$ and
define~$\mathring{\gamma}_{ab}=m_{ab}+\mathring{N}_{a}\mathring{N}_{b}$.
Let~$\phi_a$ be a spatial covector, namely~$\mathring{N}^a\phi_a=0$,
so that~$\phi_R$ and~$\phi_{S_ A}$ denote the components of~$\phi_a$
respect to the spatial frame~$( (\partial_R)^a, S_{ A}{}^a)$. To
perform the order reduction, define the reduction constraints as
\begin{align}\label{ReductionConstraintToyModel}
\mathcal{C}_a \equiv \mathring{\gamma}_a{}^{b}\mathring{\nabla}_b\phi - \phi_a,
\end{align}
and the time reduction variable~$\pi$ via~$\pi=-\partial_T\phi$. With
formally identical definitions for~$\tilde{\phi}$, the evolution
equations for these fields read,
\begin{flalign}
  \partial_T\tilde{\phi}  &=  -\tilde{\pi}, \nonumber\\
  \partial_T\tilde{\phi}_{R}{}  &=  -   \partial_R\tilde{\pi} +
  \gamma_{2}{} (- \tilde{\phi}_{R}{}
   + \partial_R\tilde{\phi}), \nonumber\\
   \partial_T\tilde{\phi}_{S_ A}{}  &= 
   \gamma_{2}{} (- \tilde{\phi}_{S_ A}{} + S_{ A}{}^{a}
  \partial_{a}\tilde{\phi})
  - S_{ A}{}^{b} \partial_{b}\tilde{\pi}, \nonumber\\
  \partial_T\tilde{\pi}  &=  - \delta^{ A B} S_{ A}{}^{a}
  \partial_{a}\tilde{\phi}_{S_{ B}}{}  - \partial_R\tilde{\phi}_{R}{}
  - \tfrac{2}{R} \tilde{\phi}_{R}
   -\tilde{\phi}_{S_ A}\mathring{\nabla}^{a}S^{ A}{}_{a},
  \nonumber\\
  \partial_T \phi  &=  - \pi, \nonumber\\
  \partial_T \phi_{R} &=  - \partial_R\pi + \gamma_{2}{} (-\phi_{R}{} +
  \partial_R\phi), \nonumber\\
  \partial_T\phi_{S_ A}{}  &=  \gamma_{2}{}
  (- \phi_{S_ A}{} + S_{ A}{}^{a}  \partial_{a}\phi)
  -  S_{ A}{}^{b} \partial_{b}\pi, \nonumber \\
     \partial_T \pi   &=  \tilde{\pi}^2 
   -  \delta^{ A B} S_{ A}{}^{a} \partial_{a} \phi_{S_{ B}}{}
   {}  -  \partial_R \phi_{R}{} - \tfrac{2}{R} \phi_{R} 
   \nonumber \\ &
   - \phi_{S_ A}\mathring{\nabla}^{a}S^{ A}{}_{a}.\label{tildenotildeeqs}
\end{flalign}
The evolution equations for $\phi$ and $\tilde{\phi}$
in~\eqref{tildenotildeeqs} are just the definition of the time
reduction variables. The evolution equations for $\phi_R$, $\phi_{S_A}$
and the corresponding hatted variables in~\eqref{tildenotildeeqs}
with~$\gamma_2=0$ arise as a consequence of the no-torsion condition
 $[\mathring{\nabla}_a,\mathring{\nabla}_b]\phi = 0$ and
$[\mathring{\nabla}_a,\mathring{\nabla}_b]\tilde{\phi} =
0$. Nevertheless, to reduce the effect produced by constraint
violations in numerical applications, one modifies these equations by
adding the following multiples of the reduction
constraints:~$\gamma_2\mathcal{C}_{S_ A}$, $\gamma_2\mathcal{C}_R$,
$\gamma_2\tilde{\mathcal{C}}_{S_ A}$,
and~$\gamma_2\tilde{\mathcal{C}}_R$. Here~$\gamma_2$ is a freely
prescribable scalar function of the coordinates. Notice that the
introduction of these terms affect the principal part of the equation.
Nonetheless, one can show that this system is symmetric hyperbolic for
any choice of the formulation
parameter~$\gamma_2$~\cite{LinSchKid05}. The evolution
equations for $\pi$ and $\tilde{\pi}$ in~\eqref{tildenotildeeqs}
arise from expressing the wave equations~\eqref{ToyModelIII}
using the reduction variables.

Following the discussion of the asymptotic system of
section~\ref{SecondOrderAsymptotic}, one needs to rescale the
variables appropriately. Nevertheless, since in the first order
reduction of the equations the frame~$((\partial_T)^a, (\partial_R),
S_{ A}{}^a)$ was used to express the components of the reduction
variable instead of the flat null frame, one needs to perform the
following change of variables before rescaling.  Defining
\begin{flalign}
& \sigma^+= -\pi+\phi_R, \quad \sigma^-=-\pi-\phi_R,
\end{flalign}
with formally identical definitions for the hatted variables,
substituting~$\pi$, $\phi_R$, $\tilde{\pi}$ and~$\tilde{\phi}_R$
written in terms of~$\sigma^+$, $\sigma^-$, $\tilde{\sigma}^+$
and~$\tilde{\sigma}^-$ into
equation~\eqref{tildenotildeeqs} one obtains a set of
evolution equations for the variables,
\begin{align}
\{ \phi, \quad \sigma^+, \quad \sigma^-, \quad \phi_{S_{ A}}, \quad
\tilde{\phi}, \quad \tilde{\sigma}^+, \quad \tilde{\sigma}^-, \quad
\tilde{\phi}_{S_{ A}} \}.\nonumber
\end{align}
Observe that when the constraints are satisfied~$\sigma^+$
and~$\sigma^-$ correspond to the~$L$ and~$\underline{L}$ derivatives
of~$\phi$, respectively. Then, one defines the \emph{rescaled
  variables} as,
\begin{align}
  \Phi & = R \phi, & \Sigma^+ &= R^2\sigma^+, \nonumber \\
  \Sigma^- & =  R\sigma^- , & \Phi_{S_ A} & = R^2\phi_{S_ A},
\end{align}
along with the analogous expressions for the hatted variables, and
substitutes these definitions into the evolution equations using the
chain rule to express the derivatives in terms of~$\partial_q$
and~$\partial_s$. Notice the important point that the rescaling here
takes place {\it after} the derivative is applied. It is easily seen
that this is the natural construction by considering spherically
symmetric solutions to the flat-space wave equation. Solving for the
derivatives of the rescaled variables we obtain,
\begin{flalign}
  \partial_q \tilde{\Sigma}^{+}{} &\simeq  \tfrac{1}{2}
  \tilde{\Sigma}^{-}{} + \tfrac{1}{2} \gamma_{2}{}
  (\tilde{\Sigma}^{+}{} + \tilde{\Phi}
   - \partial_s \tilde{\Phi}),
  \nonumber \\ \partial_s \tilde{\Sigma}^{-}{} &\simeq 
  \gamma_{2}{}{} (\tilde{\Sigma}^{+}{} + \tilde{\Phi} -
  \partial_s \tilde{\Phi}), 
   \nonumber \\ \partial_q\tilde{\Phi}_{S_ A}{}
  & \simeq - \tfrac{1}{2} \omega_{ A}{}^{a}
  \partial_{a}\tilde{\Sigma}^{-}{} + \gamma_{2}{}{}
  (\tilde{\Phi}_{S_ A}{} - \omega_{ A}{}^{a}
  \partial_{a}\tilde{\Phi}), 
  \nonumber \\  \partial_{q}\tilde{\Phi} &\simeq - \tfrac{1}{2}
  \tilde{\Sigma}^{-}{}, 
 \nonumber \\
 \partial_q\Sigma^{+}{} &\simeq \tfrac{1}{8} (4 \Sigma^{-}{} +
  (\tilde{\Sigma}^{-}){}^2) + \tfrac{1}{8} \gamma_{2}{} (4
  \Sigma^{+}{} + 4 \Phi - 4 \partial_s \Phi) ,
\nonumber  \\ \partial_s\Sigma^{-}{} &\simeq - \tfrac{1}{4}
  (\tilde{\Sigma}^{-}){}^2 + \gamma_{2}{} (\Sigma^{+}{} + \Phi -
  \partial_s \Phi),
  \nonumber \\  \partial_q\Phi_{S_ A}{} &\simeq - \tfrac{1}{2}
  \omega_{ A}{}^{a} \partial_{a}\Sigma^{-}{} + \gamma_{2}{}
  (\Phi_{S_ A}{} - \omega_{ A}{}^{a} \partial_{a}\Phi), 
 \nonumber \\ \partial_q\Phi &\simeq -
  \tfrac{1}{2} \Sigma^{-}{},\label{FirstOrderAsymptoticSystemToyBrute}
\end{flalign}
where~$\simeq$ represents equality up to error terms which decay one
order faster in $R$ than the displayed expressions. In this case, it
represents equality up to order~$O(R^{-1})$. These
expressions are derived in full in the mathematica notebooks
associated with this paper~\cite{GasHil18_web}. They require
xAct~\cite{xAct_web_aastex}.

Neglecting the error terms implicit in the last equations, namely,
formally replacing~$\simeq$ with~$=$ defines the asymptotic system for
equation~\eqref{tildenotildeeqs}. To see that this
actually corresponds to the asymptotic
system~\eqref{AsymptoticToySecondOrder} one has to examine the
relation between the rescaled variables and the corresponding rescaled
reduction constraints. Let,
\begin{flalign}
  \mathbb{C}_{R}=R^2\mathcal{C}_R, \quad\quad
  \mathbb{C}_{S_ A}=R^2\mathcal{C}_{S_ A},
\end{flalign}
with analogous definitions for the hatted reduction constraints. A
direct computation using the first and fifth equation in
\eqref{tildenotildeeqs}, equation~\eqref{ReductionConstraintToyModel},
and the definitions for the rescaled variables give,
\begin{flalign}
  \Sigma^{+}{} = & - \mathbb{C}_{R}{} - \Phi +
  \partial_s\Phi, \nonumber\\ \Sigma^{-}{} = & - 2 \partial_q\Phi
  +\frac{\mathbb{C}_{R}{}}{R} + \frac{\Phi}{R} -
  \ \frac{\partial_s\Phi}{R}, \nonumber \\ \Phi_{S_ A}{} = & -
  \mathbb{C}_{S_ A}{} + \omega_ A{}^a \partial_a\Phi,
  \label{ExpressionForUndoingTheReduction}
\end{flalign}
and identical expressions for the hatted rescaled variables.
Substituting~\eqref{ExpressionForUndoingTheReduction} and the
corresponding hatted version of these expressions into
equation \eqref{FirstOrderAsymptoticSystemToyBrute} and rearranging
one obtains,
\begin{align}
  2 \partial_{s}\partial_q \Phi &\simeq \tfrac{1}{4}
  (\tilde{\Sigma}^{-})^2 + \gamma_2 \mathbb{C}_R, & 2
  \partial_{s}\partial_q \tilde{\Phi} &\simeq \tilde{\gamma}_2
  \tilde{\mathbb{C}}_R, \nonumber \\
  \partial_q\Phi &\simeq - \tfrac{1}{2}
  \Sigma^{-}{} , & \partial_q\tilde{\Phi} &\simeq - \tfrac{1}{2}
  \tilde{\Sigma}^{-}{}, \nonumber \\
  \partial_q\mathbb{C}_{R}{} &\simeq
  \gamma_2\mathbb{C}_{R}{}, & \partial_q\tilde{\mathbb{C}}_{R}{}
  &\simeq \gamma_2 \tilde{\mathbb{C}}_{R}{}, \nonumber
  \\
  \partial_q\mathbb{C}_{S_ A}{} &\simeq \gamma_{2}{}
  \mathbb{C}_{S_ A}{}, & \partial_q\tilde{\mathbb{C}}_{S_ A}{}
  &\simeq \gamma_2 \tilde{\mathbb{C}}_{S_ A}{}.
\end{align}
These expressions are valid for any choice of the damping
parameter~$\gamma_2$. Observe that if this parameter is set to zero
or, more generally are chosen to decay sufficiently fast,
say~$\gamma_2 \simeq R^{-1}$, then one has
\begin{align}
  2 \partial_{s}\partial_q \Phi \simeq & \; \tfrac{1}{4}
  (\tilde{\Sigma}^{-})^2, & 2 \partial_{s}\partial_q \tilde{\Phi}
  \simeq &\; 0, \nonumber
  \\ \partial_q\Phi + \tfrac{1}{2} \Sigma^{-}{} \simeq & \; 0 , &
  \partial_q\tilde{\Phi} + \tfrac{1}{2} \tilde{\Sigma}^{-}{}\simeq &
  \; 0,\nonumber
  \\ \partial_q\mathbb{C}_{R}{} \simeq & \; 0, &
  \partial_q\tilde{\mathbb{C}}_{R}{} \simeq & \;
  0,\nonumber
  \\ \partial_q\mathbb{C}_{S_ A}{} \simeq &\; 0, &
  \partial_q\tilde{\mathbb{C}}_{S_ A}{} \simeq & \;
  0. \label{AsymptoticSystemToyModelUndoingTheReduction}
\end{align}
Following the philosophy of the asymptotic system by formally
replacing~$\simeq$ with~$=$ one realizes that the last four
expressions in~\eqref{AsymptoticSystemToyModelUndoingTheReduction} can
be regarded as the asymptotic equations for the rescaled reduction
constraints. Integrating these equations, along an integral curve
of~$\partial_q$, from a fixed~$q_\star$ to~$q$ reveals,
\begin{align}
\mathbb{C}_{R}{} =\mathbb{C}^\star_{R} , \quad \mathbb{C}_{S_ A}{}
= \mathbb{C}^\star_{S_ A}{}.
\end{align}
where~$\mathbb{C}^\star_{R} = \mathbb{C}_{R}|_{q=q_\star},
\mathbb{C}^\star_{S_ A}=\mathbb{C}_{S_ A}|_{q=q_\star}$ and
similarly for the hatted rescaled reduction constraints. On the other
hand, the first four equations
in~\eqref{AsymptoticSystemToyModelUndoingTheReduction} simply encode
\begin{align}
\partial_s\partial_q \Phi&= (\partial_q \tilde{\Phi})^2, \qquad
\partial_s\partial_q \tilde{\Phi}=0. 
\end{align}
Integrating these equations, this time along an integral
  curve of~$\partial_s$, we find,
\begin{align}
\Phi &=  s \; \mathcal{G}_\Phi(q,\theta^A)  ,
\quad \tilde{\Phi} = \mathcal{G}_{\tilde{\Phi}}(q,\theta^A),
\end{align}
where~$\mathcal{G}_\Phi$ and~$\mathcal{G}_{\tilde{\Phi}}$ are regular
functions of their arguments. Since we are only interested in the
behavior of the fields for large $R$, to have a more compact
notation, the symbol $\sim$ will be used to denote equality where the
functional dependence on the other coordinates has been
suppressed. Consistent with this notation one writes,
\begin{align}
 \Phi &\sim \ln R , \qquad &\tilde{\Phi} \sim 1.
\end{align}
Assuming~$\mathbb{C}_R^\star$, $\mathbb{C}^\star_{S_ A}{}$,
$\tilde{\mathbb{C}}_R^\star$ and $\tilde{\mathbb{C}}^\star_{S_ A}$
are uniformly bounded, then, using
equations~\eqref{ExpressionForUndoingTheReduction} one has that,
\begin{align}
  \Sigma^+ &\sim \ln R, & \Sigma^- &\sim \ln R, & \Phi_{S_ A} &\sim
  \ln R, \nonumber \\ \tilde{\Sigma}^+ &\sim 1, & \tilde{\Sigma}^-
  &\sim 1, & \tilde{\Phi}_{S_ A} &\sim 1.
\end{align}
Therefore one concludes that,
\begin{align}
\sigma^+ &\sim  \frac{\ln R}{R^2}, &
\sigma^- &\sim  \frac{\ln R}{R}, &
\phi_{S_ A} &\sim \frac{\ln R}{R^2}, & \nonumber \\
\tilde{\sigma}^+ &\sim \frac{1}{R^2}, &
\tilde{\sigma}^- &\sim \frac{1}{R}, &
\tilde{\phi}_{S_ A} &\sim \frac{1}{R^2}.
\end{align}
Clearly if we wished to perform numerical evolution of this system on
a compactified domain including null-infinity this result tells us
that the rescaling of variables resulting in~$\Sigma^-$ would be too
aggressive, since we need that the evolved variables are at least not
divergent.

%%%%%%%%%%%%%%%%%%%%%%%%%%%%%%%%%%%%%%%%%%%%%%%%%%%%%%%%%%%%%%%%%%%%%%%%%%%%%%%%%%%%%%%
\section{A constraint damped hyperbolic reduction of GR in second order form}
%%%%%%%%%%%%%%%%%%%%%%%%%%%%%%%%%%%%%%%%%%%%%%%%%%%%%%%%%%%%%%%%%%%%%%%%%%%%%%%%%%%%%%%
 
%%%%%%%%%%%%%%%%%%%%%%%%%%%%%%%%%%%%%%%%%%%%%%%%%%%%%%%%%%%%%%%%%%%%%%%%%%%%%%%%%%%%%%%
\subsection{Hyperbolic reductions of GR}
%%%%%%%%%%%%%%%%%%%%%%%%%%%%%%%%%%%%%%%%%%%%%%%%%%%%%%%%%%%%%%%%%%%%%%%%%%%%%%%%%%%%%%%

Let~$(\mathcal{M},g_{ab})$ denote a 4-dimensional manifold equipped
with a metric~$g_{ab}$. Let~$m_{ab}$ denote the Minkowski metric and
let~$\nabla$ and~$\mathring{\nabla}$ denote the Levi-Civita connection
of~$g_{ab}$ and~$m_{ab}$, respectively. The relation between~$\nabla$
and~$\mathring{\nabla}$ can be parameterized via,
\begin{align}
\nabla_av_b = \mathring{\nabla}_av_b - \Gamma^{c}{}_{ab} v_c,
\end{align}
where~$v_a$ is any covector. This relation can be taken to
define~$\Gamma^{c}{}_{ab}$. Consequently~$\Gamma^{c}{}_{ab}$ can be
expressed in terms of~$\mathring{\nabla}$-derivatives of~$g_{ab}$ as
\begin{align}\label{eq:Koszul}
\Gamma^{c}{}_{ab}&=\tfrac{1}{2} g^{cd} (\mathring{\nabla}_{a}g_{bd}
+ \mathring{\nabla}_{b}g_{ad} - \mathring{\nabla}_{d}g_{ab}).
\end{align}
Defining the contracted Christoffel symbols via~$\Gamma^c \equiv
g^{ab}\Gamma^{c}{}_{ab}$, the Ricci tensor can be compactly expressed
as,
\begin{align}
 R_{ab} &= - \tfrac{1}{2} g^{cd}
 \mathring{\nabla}_{c}\mathring{\nabla}_{d}g_{ab} +
 \nabla_{(a}\Gamma_{b)} \nonumber\\
 &\quad+ g^{cd} g^{hf} (\Gamma_{bdf}\Gamma_{cah} +
 2\Gamma_{dbf}\Gamma_{(ac)h}).
 \label{eq:RicciSecondDerivativesMetric}
\end{align}
Let~$C^a \equiv F^a + \Gamma^a$ where~$F^a$ are smooth functions of
the coordinates~$X^\mu$ and the metric~$g_{ab}$ but not its
derivatives.  These are known as the gauge source functions as a
choice of $F^a$ determines a coordinate system $X^\mu$. To see this,
observe that $\Gamma^\mu=-\nabla^\nu\nabla_\nu X^\mu$, thus requiring
$C^\mu=0$ is equivalent to
\begin{align}
\nabla^\nu\nabla_\nu X^\mu=F^\mu.
\end{align}
The equations $C^a=0$ will be called GHG or 
harmonic constraints in the $F^a=0$ case. If
they are satisfied then, using
equation~\eqref{eq:RicciSecondDerivativesMetric}, one sees that the
vacuum Einstein field equations reduce to a set of wave equations
for~$g_{ab}$
\begin{align}
   g^{cd}
  \mathring{\nabla}_{c}\mathring{\nabla}_{d}g_{ab}{}
  &= 2 g^{cd} g^{hf}
  (\Gamma_{bdf}\Gamma_{cah} 
  + 2\Gamma_{dbf}\Gamma_{(ac)h}){}\nonumber\\
  &\quad+2\nabla_{(a}F_{b)},\label{eq:reducedEFEversion1}
\end{align}
where~$\Gamma_{abc}$ is expressed in terms of derivatives of the
metric~$g_{ab}$ using equation~\eqref{eq:Koszul}. This hyperbolic
reduction process can be succinctly expressed as follows. Define the
reduced Ricci tensor as
\begin{align}
\underline{R}_{ab}=R_{ab}-\nabla_{(a}C_{b)},
\end{align}
so that equation~\eqref{eq:reducedEFEversion1} is encoded in the
condition~$\underline{R}_{ab}=0$. Observe that if the
constraint~$C^a=0$ is satisfied
then~$\underline{R}_{ab}=R_{ab}$. Moreover, if~$\underline{R}_{ab}=0$
then, as a consequence of the contracted second Bianchi
identity~$\nabla^aR_{ab}-\frac{1}{2}\nabla_bR=0$, one has that~$C_a$
satisfies the following propagation equation
\begin{align}
\nabla^a\nabla_aC_{b} = - C^a\nabla_{(a}C_{b)}.
\end{align}
Since this is a wave equation homogeneous in~$C_a$, the latter implies
that if~$C_{a}$ and~$\nabla_aC_{b}$ vanish on a spacelike
hypersurface~$\mathcal{S}\subset\mathcal{M}$ then~$C_a=0$
in~$\mathcal{D}(\mathcal{S})\subset\mathcal{M}$~\cite{Tay96,GomVal08}. Here
the domain of dependence of an achronal set~$\mathcal{A}$, is denoted
as~$\mathcal{D}(\mathcal{A})$. Observe that this hyperbolic reduction
strategy is not unique as one can define a reduced Ricci tensor as
\begin{align}\label{ReducedRicciTensorDefinitionGeneral}
\underline{R}_{ab}=R_{ab}-\nabla_{(a}C_{b)} + \mathcal{T}_{ab}.
\end{align}
where~$\mathcal{T}_{ab}$ is any expression homogeneous in~$C_a$ so
that~$C_a=0$ implies~$\underline{R}_{ab}=R_{ab}$. The corresponding
propagation equation for~$C_a$ is then,
\begin{flalign}
 \nabla^{a}\nabla_{a}C_{b} = - C^{a} \nabla_{(a}C_{b)} + 2
\nabla^{a}\mathcal{T}^{\text{TR}}_{ab}+
C^{a} \mathcal{T}_{ba}, 
\end{flalign}
where $\mathcal{T}^{\text{TR}}_{ab} \equiv
\mathcal{T}_{ab}-\tfrac{1}{2}g_{ab}\mathcal{T}_{c}{}^{c}$.  Observe
that, the right hand side of the last equation is homogeneous
in~$C_{a}$ as long as~$\mathcal{T}_{ab}$ is chosen to be homogeneous
in~$C_{a}$. Although all the possible reduced
equations~$\underline{R}_{ab}=0$ are equivalent if the GHG constraints
are satisfied, a different choice of~$\mathcal{T}_{ab}$ can be used to
obtain equations of a particular desired form. For instance
in~\cite{GunGarCal05}, the constraint addition is chosen
as~$\mathcal{T}_{ab} = {}^{\gamma}\mathcal{T}_{ab}$ where,
\begin{align}
{}^{\gamma}\mathcal{T}_{ab}&=\gamma_4\Gamma^c{}_{ab}C_{c}-\frac{1}{2}\gamma_5
g_{ab}\Gamma^eC_e -\gamma_0(n_{(a}C_{b)}-g_{ab}n^cC_{c}),
\end{align}
here~$n_a$ is a freely specifiable vector
and~$\gamma_0,\gamma_4,\gamma_5$ are, in general, scalar functions
depending on the coordinates. The parameter~$\gamma_0$ is included to
damp away high frequency constraint violations while the
parameters~$\gamma_4$ and~$\gamma_5$ are included to
modify~$\nabla_{(a}C_{b)}$ so the constraint addition made in the
construction of the formulation is done either respect to~$\nabla$
or~$\mathring{\nabla}$, or some combination thereof.

Other choices are designed instead to exhibit certain structures of
the equations. In particular, if one sets~$\mathcal{T}_{ab}
={}^{\mathcal{C}}\mathcal{T}_{ab}$ where
\begin{align}\label{eq:LinRodConstraintAddition}
{}^{\mathcal{C}}\mathcal{T}_{ab}=\frac{1}{2}C_{a}C_{b}-C^c\mathring{\nabla}_{(a}g_{b)c}.
\end{align}
Then the reduced Einstein field equations $\underline{R}_{ab}=0$ read
\begin{align}\label{eq:reducedEFELindblad}
  g^{cd}\mathring{\nabla}_{d}\mathring{\nabla}_{c}g_{ab}=P_{ab}[\mathring{\nabla}g]
  +Q_{ab}[\mathring{\nabla}g]+F_{ab},
\end{align}
with
\begin{align}
  P_{ab}[\mathring{\nabla} g] &
  =- \tfrac{1}{2} g^{cd} g^{fh} \mathring{\nabla}_{a}g_{cf}
  \mathring{\nabla}_{b}g_{dh} %\nonumber  %\\ & \quad
  + \tfrac{1}{4} g^{cd} g^{fh}
\mathring{\nabla}_{a}g_{cd} \mathring{\nabla}_{b}g_{fh},\label{eq:pnotnullform}
\\ 
Q_{ab}[\mathring{\nabla}g] & = -g^{cd} g^{fh}\big( \tfrac{1}{2}
\mathring{\nabla}_{b}g_{fh} \mathring{\nabla}_{d}g_{ac} - \tfrac{1}{2}
\mathring{\nabla}_{a}g_{fh} \mathring{\nabla}_{d}g_{bc}\nonumber \\ &
+ {} \tfrac{1}{2}{}
\mathring{\nabla}_{a}g_{bc} \mathring{\nabla}_{d}g_{fh}{}
+{} \tfrac{1}{2} \mathring{\nabla}_{b}g_{ac}
\mathring{\nabla}_{d}g_{fh} {}+{}
\mathring{\nabla}_{b}g_{dh}\mathring{\nabla}_{f}g_{ac}\nonumber \\ & -
\mathring{\nabla}_{d}g_{bh} \mathring{\nabla}_{f}g_{ac} +
\mathring{\nabla}_{a}g_{dh} \mathring{\nabla}_{f}g_{bc}
+\mathring{\nabla}_{f}g_{ac} \mathring{\nabla}_{h}g_{bd} \nonumber
\\ & + \mathring{\nabla}_{d}g_{ac} \mathring{\nabla}_{h}g_{bf} -
\mathring{\nabla}_{a}g_{bc} \mathring{\nabla}_{h}g_{df} -
\mathring{\nabla}_{b}g_{ac} \mathring{\nabla}_{h}g_{df} \big)\label{eq:qnullform},
\end{align} 
and the gauge source functions~$F^a$ appear in the form,
\begin{flalign}
F_{ab} = & \; 2\nabla_{(a}F_{b)} - F_{a}F_{b} + 2F^c\mathring{\nabla}_{(a}g_{b)c}.
\end{flalign}
The relevant observation here is that~$Q_{ab}$ is a null
form. Therefore, taking~$F^a=0$, the constraint
addition~\eqref{eq:LinRodConstraintAddition} places the reduced field
equations~$\underline{R}_{ab}=0$ in the form given in~\cite{LinRod05}
where the global stability of the Minkowski spacetime in wave
coordinates was proven for a set of restricted initial data, when the
harmonic constraints are fulfilled.

%%%%%%%%%%%%%%%%%%%%%%%%%%%%%%%%%%%%%%%%%%%%%%%%%%%%%%%%%%%%%%%%%%%%%%%%%%%%%%%%%%%%%%%
\subsection{The asymptotic system and the weak null condition}
\label{sec:AsymptoticSystemLindRod}
%%%%%%%%%%%%%%%%%%%%%%%%%%%%%%%%%%%%%%%%%%%%%%%%%%%%%%%%%%%%%%%%%%%%%%%%%%%%%%%%%%%%%%%
 
In this section the asymptotic system for the Einstein field equations
in harmonic gauge as discussed in~\cite{LinRod03,LinRod05,Lin17} is
reviewed in our conventions and the problems that the above hyperbolic
reductions present for free evolution schemes in NR are discussed.

Consider a perturbation of the Minkowski spacetime,
and write $g_{ab}$ as
\begin{align}\label{eq:splitg}
  g_{ab}= m_{ab} + h_{ab},
\end{align}
where~$h_{ab}$ is a symmetric 2-tensor. For the discussion in the
remainder of this subsection all the indices are moved using the flat
metric $m_{ab}$ except for $g_{ab}$ for which we have
\begin{align}\label{eq:upsplitg}
g^{ab}=m^{ab}-h^{ab}+O^{ab}(h^2),
\end{align}
where~$h^{ab}=m^{ac}m^{db}h_{cd}$ and $O^{ab}(h^2)$ vanishes to second
order at~$h=0$~\cite{LinRod04}. This
convention for raising and lowering indices will be used when discussing
the derivation of the asymptotic system of a given set
of equations but it will be avoided otherwise. Now, consider the
reduced Ricci operator~$\underline{R}_{ab}=R_{ab}-\nabla_{(a}C_{b)} +
\mathcal{T}_{ab}$
with~$\mathcal{T}_{ab}={}^{\mathcal{C}}\mathcal{T}_{ab}$ as in
equation~\eqref{eq:LinRodConstraintAddition} in harmonic gauge, in
other words, with vanishing gauge source functions~$F^a=0$. In this
case, the reduced Einstein field equations~$\underline{R}_{ab}=0$
imply the following system of wave equations for~$g_{ab}$
\begin{align}\label{eq:ReducedEFELindRod}
g^{cd}  \mathring{\nabla}_{d}\mathring{\nabla}_{c}g_{ab}=P_{ab}[\mathring{\nabla}g]
  +Q_{ab}[\mathring{\nabla}g].
\end{align}
Using equation~\eqref{eq:splitg} and~\eqref{eq:upsplitg} one obtains
the following evolution equations for the {\it perturbation}~$h_{ab}$
\begin{align}\label{eq:eveqpertHarmonic}
g^{cd}\mathring{\nabla}_c\mathring{\nabla}_d h_{ab} =
  P_{ab}[\mathring{\nabla}h]+Q_{ab}[\mathring{\nabla}h]
  + O_{ab}(h(\mathring{\nabla}h)^2),
\end{align}
with~$P_{ab}[\mathring{\nabla}h]$ and~$Q_{ab}[\mathring{\nabla}h]$ as
given in equations~\eqref{eq:pnotnullform}-\eqref{eq:qnullform}
where~$\mathring{\nabla} g$ is formally replaced
by~$\mathring{\nabla}h$ and~$O_{ab}(h(\mathring{\nabla}h)^2)$ denotes
a quadratic form in $\mathring{\nabla}_ah_{bc}$ with coefficients
depending on~$h_{ab}$ which vanish for~$h_{ab}=0$.  Similarly, the
harmonic coordinate condition reads
\begin{align}
m^{ac}\mathring{\nabla}_c
h_{ab}=\frac{1}{2}m^{ac}\mathring{\nabla}_{b}h_{ac} +
O_{b}(h\mathring{\nabla}h),
\end{align}
where~$O_{b}(h\mathring{\nabla}h)$ is linear
in~$\mathring{\nabla}_ah_{bc}$ with coefficients depending on $h_{ab}$
which vanish for $h_{ab}=0$.  To derive the asymptotic system, let
\begin{align}\label{eq:AsymptSystemRecipe}
q=R-T, \quad s=\ln R, \quad H_{ab}=R \; h_{ab}.
\end{align}
To have a more compact notation for directional derivatives that will
often appear, the following notation will be used,
\begin{flalign}
\mathring{\nabla}_T &= (\partial_T)^a\mathring{\nabla}_a, \qquad \nonumber
\mathring{\nabla}_R = (\partial_R)^a\mathring{\nabla}_a,\\
\mathring{\nabla}_q &= (\partial_q)^a\mathring{\nabla}_a, \qquad 
\mathring{\nabla}_s = (\partial_s)^a\mathring{\nabla}_a.
\end{flalign}
Substituting~\eqref{eq:splitg} into~\eqref{eq:reducedEFELindblad},
re-expressing the equation in terms of~$H_{ab}$ and formally equating
terms with coefficients~$R^{-2}$, one obtains the following asymptotic
system
\begin{align}\label{eq:AsymptEFELindRod}
(2 \mathring{\nabla}_{s} - H_{LL}{}
\mathring{\nabla}_{q})\mathring{\nabla}_{q}H_{ab} = \hat{L}_a\hat{L}_b
P(\mathring{\nabla}_{q}H,\mathring{\nabla}_{q}H),
\end{align}
where~$H_{LL}=L^{a}L^{b}H_{ab}$ and
\begin{flalign}
P(\mathring{\nabla}_{q}H) = & - \frac{1}{2}m^{ab} m^{cd}
\mathring{\nabla}_{q} H_{ac} \
\mathring{\nabla}_{q}H_{bd} \nonumber \\ & + \frac{1}{4} m^{ab} m^{cd}
\ \mathring{\nabla}_{q}H_{ab}
\mathring{\nabla}_{q}H_{cd}.
\end{flalign}
Contracting equations~\eqref{eq:AsymptEFELindRod} with the flat null
frame~$(L^{a},\underline{L}^a, S_{ A}{}^a)$, exploiting that
\begin{align}
v^aw^b\mathring{\nabla}_cH_{ab} & \simeq \mathring{\nabla}_cH_{vw}, \nonumber \\
v^aw^b\mathring{\nabla}_s\mathring{\nabla}_qH_{ab} &
\simeq {\partial}_s{\partial}_qH_{vw}, \nonumber \\
v^aw^b\mathring{\nabla}_q\mathring{\nabla}_qH_{ab} &
\simeq {\partial}_q{\partial}_qH_{vw},
\end{align}
for~$v^a, w^a \in \{L^a,\underline{L}^a,S_{ A}{}^a\}$, one observes
that the only equation in~\eqref{eq:AsymptEFELindRod} with a
non-vanishing right-hand side is,
\begin{flalign}\label{eq:BadComponentPerturbation}
(2 \partial_{s} - H_{LL}{}
\partial_{q})\partial_{q}H_{\underline{L}\underline{L}} &= 
-4(\partial_qH^{+})^2 -4 (\partial_qH^{\times})^2 \nonumber \\
& - \partial_{q}H_{\underline{L} \underline{L}}{} \partial_{q}H_{LL}{} 
- 2 (\partial_q H_{L\underline{L}} )(\partial_q H^{\varnothing})\nonumber \\&
+ 4\delta^{ A B}(\partial_q H_{LS_{ A}})(\partial_q H_{\underline{L}S_{ B}}),
\end{flalign}
while all the others satisfy,
\begin{align}\label{eq:GoodComponentsPerturbation}
(2 \partial_{s} - H_{LL}{} \partial_{q})\partial_{q}H_{\mathcal{T}U}=0,
\end{align}
where~$\mathcal{T} \in \{L, S_{ A}\}$ and~$U \in \{L,\underline{L},S_{ A}\}$.
Proceeding similarly for the harmonic coordinate
condition
\begin{equation}
  m_{ab}C^{b}=m_{ab}\Gamma^{b}=m_{ab}F^{b}=0,
\end{equation}
using equations~\eqref{eq:splitg}
and~\eqref{eq:AsymptSystemRecipe}, and formally equating terms with
coefficients~$R^{-1}$, \emph{the asymptotic harmonic coordinate
  condition} reads
\begin{align}\label{eq:AsymptHarmonicCondition}
  \mathring{\nabla}_{q}H_{L}{}_{a} - \tfrac{1}{2}
  \hat{L}_{a} \ \mathring{\nabla}_{q}H = 0,
\end{align}
where~$H_{La}=L^cH_{ca}$ and~$H=m^{ab}H_{ab}$.

Expressing equation~\eqref{eq:AsymptHarmonicCondition} in components
respect to the flat null frame one has,
\begin{align}
  & \partial_q H_{L\mathcal{T}}=0, & 
  \partial_q H^{\varnothing} &=0,\label{eq:AsymptHarmonicConditionComps}
\end{align}
where~$H^{\varnothing}$ is defined according to equation \eqref{TraceonSphere}.
The asymptotic harmonic condition~\eqref{eq:AsymptHarmonicCondition}
will play an important role in the subsequent discussion since in free
evolution schemes in NR one cannot ensure that the constraint
equations~$C^{a}=0$ are satisfied but only that such violations are
small. Moreover, notice that the asymptotic
equation~\eqref{eq:GoodComponentsPerturbation} implies
that~$\partial_{q}H_{\mathcal{T}U}$ is constant along the integral
curves of the vector field~$2 \partial_{s} - H_{LL}{}\partial_{q}$.
Observing that constraint violations~$C^a\neq 0$ imply that
$(\partial_qH_{L\mathcal{T}})|_{\Sigma}\neq0$ one concludes that, if
constraint violations are present, they will be preserved in the
asymptotic region along the integral curves of~$2 \partial_{s} -
H_{LL}{} \partial_{q}$.  As an additional remark, observe that, if the
constraints~$C^{a}=0$ are satisfied, then one can exploit
equations~\eqref{eq:AsymptHarmonicConditionComps} to reduce the
asymptotic equation~\eqref{eq:BadComponentPerturbation} to
\begin{flalign}\label{eq:SimplifiedEquationForBadComponent}
 (2 \partial_{s} - H_{LL}{}
  \partial_{q})\partial_{q}H_{\underline{L}\underline{L}} =
  -4(\partial_qH^{+})^2 -4 (\partial_qH^{\times})^2,
\end{flalign}
where~$H^{\times}$ and $H^+$ correspond to the two gravitational
wave polarizations as in equation \eqref{TracefreePlusCross}.
 This further simplification in the only
equation of the asymptotic system~\eqref{eq:AsymptEFELindRod} with
non-vanishing right-hand side cannot be directly attained in free
evolution form without modifying the hyperbolic reduction determined
by the constraint addition~$^{\mathcal{C}}\mathcal{T}_{ab}$. Thus, one
requires a hyperbolic reduction for which one recovers the latter
equation without assuming that the constraint
equations~\eqref{eq:AsymptHarmonicConditionComps} are fulfilled and,
more importantly, a hyperbolic reduction for which constraint
violations~$C^a \neq 0$ are damped close to null infinity.

%%%%%%%%%%%%%%%%%%%%%%%%%%%%%%%%%%%%%%%%%%%%%%%%%%%%%%%%%%%%%%%%%%%%%%%%%%%%%%%%%%%%%%%
\subsection{The coordinate light-speed condition}
\label{sec:coordlight-speed}
%%%%%%%%%%%%%%%%%%%%%%%%%%%%%%%%%%%%%%%%%%%%%%%%%%%%%%%%%%%%%%%%%%%%%%%%%%%%%%%%%%%%%%%

In this subsection asymptotic expressions for quantities relevant for the
dual foliation formulation are written in terms of the variables of
the asymptotic system. In particular, we are interested in finding how
the solution to the asymptotic system determines the radial coordinate
light-speed asymptotically near null-infinity. To connect the current
discussion with the dual foliation formalism, is convenient to perform
a 2+1+1 split.  Since in this subsection we are not computing the
asymptotic system we will use the metric $g_{ab}$ ---as usual--- to
raise and lower indices and not $m_{ab}$. In the language of~\cite{Hil15}
let~$(T,R,\theta^{A})$ be the \emph{upper case} coordinate system
$X^\mu$ and use~$T$ to define the usual lapse, projection operator,
normal and shift vectors
\begin{align}
\mathcal{A} \equiv & ( -g^{ab}\nabla_a T \nabla_a T )^{-1/2}, &
N_a \equiv & -\mathcal{A}\nabla_a T, \nonumber \\ \gamma_{ab} \equiv &
g_{ab}+N_{a}N_{b}, & \mathcal{B}_a \equiv& \gamma_a{}^b
\nabla_bT.
\end{align}
Similarly, we use the coordinate~$R$ to define the corresponding
normal vector~$S^a$, projector~$q_{ab}$, length scalar~$\mathcal{L}$
and slip vector~$b^a$
\begin{align}
\mathcal{L} \equiv & ( \gamma^{ab}\nabla_a R \nabla_b R )^{-1/2}, &
S_a \equiv & \mathcal{L} \gamma_{a}{}^{b}\nabla_b R, \nonumber \\ 
q_{ab}\equiv & \gamma_{ab}-S_{a}S_{b}, & b_a \equiv & q_a{}^b \nabla_bR.
\end{align}
Using the above definitions, the metric~$g_{ab}$, written as a line
element, reads~\cite{Hil15},
\begin{flalign}\label{eq:DFsplitMetric}
\mbox{d}S^2    = &
-\mathcal{A}^2\mbox{d}T^2+\mathcal{L}^2(\mbox{d}R
+\mathcal{L}^{-1}\mathcal{B}^R\mbox{d}T)^2
+ q_{AB} (\mbox{d}\theta^A \nonumber \\ &
+b^{A}\mbox{d}R+\mathcal{B}^{A}\mbox{d}T)(\mbox{d}\theta^B
+b^{B}\mbox{d}R+\mathcal{B}^{B}\mbox{d}T).
\end{flalign}
On the other hand, the split~$g_{ab}=m_{ab}+h_{ab}$ implies
\begin{flalign}\label{eq:mplushSplitSpherical}
\mbox{d}s^2 = & (-1+h_{TT})\mbox{d}T^2
+2h_{TR}\mbox{d}R\mbox{d}T
+(1+h_{RR})\mbox{d}R^2 \nonumber \\ &
+2\slashed{h}_{RA}\mbox{d}\theta^A\mbox{d}R
+2\slashed{h}_{TA}\mbox{d}\theta^A\mbox{d}T
+ (R^2\slashed{\delta}_{AB} \nonumber \\ &
+\slashed{h}_{AB})\mbox{d}\theta^A\mbox{d}\theta^B,
\end{flalign}
where
\begin{align}
\slashed{\delta}_{AB} & =
\Omega_{A}{}^{ C}\Omega_{B}{}^{ D}\delta_{ C D}, &
\slashed{h}_{AB} & =
\Omega_{A}{}^{ C}\Omega_{B}{}^{ D}h_{\omega_{ C}\omega_{ D}},
\nonumber \\ \slashed{h}_{RA} & =
\Omega_{A}{}^{ B}h_{R\omega_{ B}}, & \slashed{h}_{TA} & =
\Omega_{A}{}^{ B}h_{T\omega_{ B}}.
\end{align}
Comparing expressions~\eqref{eq:DFsplitMetric}
and~\eqref{eq:mplushSplitSpherical} gives
\begin{align}
-1+ h_{TT} &= -\mathcal{A}^2+(\mathcal{B}^{R})^2 +
q_{AB}\mathcal{B}^{A}\mathcal{B}^B,
\nonumber \\ 1+h_{RR}
&= \mathcal{L}^2 + q_{AB}b^{A}b^{B},\nonumber \\ h_{RT} &=
q_{AB}b^A\mathcal{B}^B + \mathcal{L}
\mathcal{B}^R, \nonumber \\ q_{AB} &=
R^2\slashed{\delta}_{AB}+\slashed{h}_{AB}, \nonumber
\\ \slashed{h}_{RA} &= q_{AB}b^B
, \nonumber \\ \slashed{h}_{TA} &=
q_{AB}\mathcal{B}^B. \label{eq:htAtoShift}
\end{align}
Inverting the matrix encoded in the fifth equation
in ~\eqref{eq:htAtoShift} to
write~$q^{AB}$ in terms of~$\slashed{\delta}_{AB}$
and~$\slashed{h}_{AB}$, the variables of the dual foliation
formulation can be expressed as,
\begin{flalign}
\mathcal{A} &= \bigg(
\frac{(h_{RT}-q^{AB}\slashed{h}_{RA}\slashed{h}_{TB})^2}
{h_{RR}+1-q^{AB}\slashed{h}_{RA}\slashed{h}_{RA}} \nonumber \\
& - (h_{TT}-1 - q^{AB}\slashed{h}_{TA}\slashed{h}_{TA}) \bigg)^{1/2},
\nonumber \\
\mathcal{B}^R &=\frac{h_{RT}-q^{AB}\slashed{h}_{RA}\slashed{h}_{TB}}
{\sqrt{h_{RR}+1-q^{AB}\slashed{h}_{RA}\slashed{h}_{RA}}},
\nonumber \\
\mathcal{L}&=
\sqrt{h_{RR}+1-q^{AB}\slashed{h}_{RA}\slashed{h}_{RB}}, \nonumber\\
b^A&= q^{AB}\slashed{h}_{RB}, \nonumber \\ \mathcal{B}^A &=q^{AB}\slashed{h}_{TB}.
\label{eq:DFvariablesTohSpherical}
\end{flalign}
The last expressions can be written in terms of the components
of~$h_{ab}$ in the flat null frame using that
\begin{align}
&h_{TT}=\frac{1}{4}(h_{LL}+2h_{L\underline{L}}+h_{\underline{L}\underline{L}}), \nonumber\\
&h_{RR}=\frac{1}{4}(h_{LL}-2h_{L\underline{L}}+h_{\underline{L}\underline{L}}),\nonumber \\
& h_{RT}=\frac{1}{4}(h_{LL}-h_{\underline{L}\underline{L}}),\nonumber \\
  &\slashed{h}_{RA}=\frac{1}{2}\Omega_{A}{}^{ B}(h_{L\omega_{ B}}
  -h_{\underline{L}\omega_{ B}}),\nonumber
  \\ &\slashed{h}_{TA}=\frac{1}{2}\Omega_{A}{}^{ B}
  (h_{L\omega_{ B}}+h_{\underline{L}\omega_{ B}}).\label{eq:SpCompsPertToFlatNull}
\end{align}
Recalling that~$h_{ab}=\frac{1}{R} H_{ab}$ and taking into account the
relation between~$\omega_{ A}{}^a$ and $S_{ A}{}^a$, as given in
the expressions~\eqref{eq:NullFrameToSpFrame}, gives
\begin{align}
h_{LL}&=  \frac{1}{R}H_{LL}, &
h_{L\underline{L}}&= \frac{1}{R}H_{L\underline{L}}, \nonumber \\ 
h_{\underline{L}\underline{L}}&=  \frac{1}{R}H_{\underline{L}\underline{L}},
&\slashed{h}_{LA} &= \Omega_{A}{}^{ B}H_{LS_{ B}}, \nonumber \\
\slashed{h}_{\underline{L}A} &=  \Omega_{A}{}^{ A}H_{\underline{L}S_{ A}},
&\slashed{h}_{AB} &=  R\Omega_{A}{}^{ C}\Omega_{B}{}^{ D}H_{S_{ C}S_{ D}}.
\label{eq:hnullToHnull}
\end{align}
Now, recall that the radial and angular coordinate light-speeds in the
dual foliation formulation~\cite{Hil15, HilHarBug16} are
given by
\begin{align}\label{eq:CoordinateLight-Speed}
C^{R}_\pm =-\mathcal{B}^R \pm \mathcal{A}\mathcal{L}^{-1}, \qquad
C^{A}_\pm =-\mathcal{B}^A \mp b^{A}\mathcal{A}\mathcal{L}^{-1}.
\end{align}
Thus, using equations \eqref{eq:DFvariablesTohSpherical},
\eqref{eq:SpCompsPertToFlatNull} and~\eqref{eq:hnullToHnull} we obtain
\begin{flalign}
& C_{+}^{R} \simeq 1-\frac{H_{LL}}{2R},  \qquad
C_{-}^{R} \simeq -1+\frac{H_{\underline{L}\underline{L}}}{2R},\nonumber \\
& C^{A}_\pm = O^A(R^{-2}).
\end{flalign}
For completeness of the discussion observe that the variables of the
dual foliation formulation are related to the variables of the
asymptotic system via
\begin{align}
\mathcal{A} &\simeq \pm 1 \mp
\frac{H_{\underline{L}\underline{L}}+2H_{L\underline{L}}+H_{LL}}{8R},
\nonumber \\ \mathcal{L} &\simeq \pm 1
\pm\frac{H_{\underline{L}\underline{L}}-2H_{L\underline{L}}+H_{LL}}{8R},
\nonumber \\ \mathcal{B}^{R} &\simeq
\pm\frac{H_{LL}-H_{\underline{L}\underline{L}}}{4R}, \nonumber \\ b^A
&= O^A(R^{-2}), \nonumber \\ \mathcal{B}^A &=O^A(R^{-2}), \nonumber
\\ q_{AB}
&\simeq\Omega_{A}{}^{ A}\Omega_{A}{}^{ B}(R^2\delta_{ A B}+ R
H_{S_{ A}S_{ B}}). \label{qABexpansion}
\end{align}
Using the chain rule one has that
\begin{flalign}
&\partial_{T}C_{+}^{R}\simeq\frac{1}{2R}\partial_{q}H_{LL}, \nonumber \\
& \partial_{R}C_{+}^{R}\simeq \frac{1}{2R^2}(H_{LL} -
\partial_{s}H_{LL})-\frac{1}{2R}\partial_{q}H_{LL}.
\end{flalign}
The coordinate light-speed condition~\cite{HilHarBug16} is that
\begin{align}
\partial_{T}C_{+}^{R}=\partial_{\underline{i}}C_{R}^{+}=O(R^{-1-\delta})
\end{align}
with $\delta>0$.  Consequently, the coordinate light-speed condition
is satisfied as long as
\begin{align}\label{eq:DFcondition}
H_{LL}+\partial_{s}H_{LL}=O(R^{1-\delta}), \qquad
\partial_{q}H_{LL}=O(R^{-\delta}).
\end{align}

%%%%%%%%%%%%%%%%%%%%%%%%%%%%%%%%%%%%%%%%%%%%%%%%%%%%%%%%%%%%%%%%%%%%%%%%%%%%%%%%%%%%%%%
\section{The asymptotic system
for~$\mathcal{T}_{ab}={}^{\mathscr{I}}\mathcal{T}_{ab}$}\label{sec:GasHilConstraintAddition}
%%%%%%%%%%%%%%%%%%%%%%%%%%%%%%%%%%%%%%%%%%%%%%%%%%%%%%%%%%%%%%%%%%%%%%%%%%%%%%%%%%%%%%%

In view of the remarks on made in
subsection~\ref{sec:AsymptoticSystemLindRod} about free evolution
schemes in NR, the following hyperbolic reduction of the Einstein
field equations will be considered. Let
\begin{align}\label{ConstraintAdditionTotal}
  {}^{\mathscr{I}}\mathcal{T}_{ab} \equiv
  {}^{\mathcal{C}}\mathcal{T}_{ab} + W_{ab} + \mathcal{W}_{ab}
\end{align}
here $W_{ab}$ and $\mathcal{W}_{ab}$ are symmetric tensors expressed
in the flat null frame as in
equation~\eqref{eq:SymmetricTensorInFlatNullFrame} where the only
non-vanishing components are given by
\begin{flalign}
  W_{\underline{L}\underline{L}}  &=  
  C^c\mathring{\nabla}_{\underline{L}}g_{\underline{L}c}-
  \tfrac{1}{2}C^c\mathring{\nabla}_cg_{\underline{L}\underline{L}} \nonumber \\
  \mathcal{W}_{LL} &=  \tfrac{1}{2}
  C_{L}C_{L} -\tfrac{ \omega}{R} C_{L},
  \nonumber \\
  \mathcal{W}_{LS_{ A}}&= \tfrac{1}{2} C_{L}C_{S_{ A}} -
  \tfrac{\omega}{R} C_{S_{ A}}, \nonumber \\
  \mathcal{W}_{S_{ A}S_{ B}}&= \tfrac{1}{2}\delta_{ A B}\Big( \tfrac{1}{2} 
   C_{L}C_{\underline{L}}  -  \tfrac{ \omega}{R} C_{\underline{L}}\Big).
\end{flalign}
In these expressions~$\omega$ is a positive constant which will play a
similar role to the damping parameters~$\gamma_0,\gamma_4$ and
$\gamma_5$ in~\cite{GunGarCal05}. First notice
that~${}^{\mathscr{I}}\mathcal{T}_{ab}$ is homogeneous in~$C_a$ so
that~$C_a=0$ implies~${}^{\mathscr{I}}\mathcal{T}_{ab}=0$.

The reduced Einstein field equations~$\underline{R}_{ab}=0$ imply the
following set of wave equations,
\begin{align}
g^{cd} \mathring{\nabla}_{d}\mathring{\nabla}_{c}g_{ab}
&=P_{ab}[g]+Q_{ab}[g]+F_{ab} + 2W_{ab} +2 \mathcal{W}_{ab},
\label{eq:ReducedEFEGasHil}
\end{align}
for the metric. Now, we consider $g_{ab} = m_{ab} + h_{ab}$ to
compute the corresponding evolution equations for $h_{ab}$ and then
its asymptotic system.  Here, since we are to discuss the derivation
of the corresponding asymptotic system, it is understood, as in
subsection \ref{sec:AsymptoticSystemLindRod}, that the metric $m_{ab}$ is
used to raise and lower indices of all tensors except for $g_{ab}$ for
which one uses equation \eqref{eq:upsplitg}.  Taking these
considerations into account, a direct computation using equation
\eqref{eq:ReducedEFEGasHil} renders the following evolution equations
for ~$h_{ab}$,
\begin{align}
g^{cd}\mathring{\nabla}_c\mathring{\nabla}_d h_{ab} &=
P_{ab}[h]+Q_{ab}[h]+F_{ab}  \nonumber\\
&+ 2W_{ab} +2
\mathcal{W}_{ab} + O_{ab}(h(\mathring{\nabla}h)^2),
\end{align}
where it is understood that~$F_{ab}$, $W_{ab}$ and~$\mathcal{W}_{ab}$
had been expanded up to
order~$O_{ab}(h(\mathring{\nabla}h)^2)$. For completeness
observe that the GHG constraint reads
\begin{align}
m^{ac}\mathring{\nabla}_c
h_{ab}=\frac{1}{2}m^{ac}\mathring{\nabla}_{b}h_{ac} -F_b +
O_{b}(h\mathring{\nabla}h),
\end{align}
where~$O_{b}(h\mathring{\nabla}h)$ is linear
in~$\mathring{\nabla}_ah_{bc}$ with coefficients depending on~$h_{ab}$
which vanish for~$h_{ab}=0$.

Recall that the asymptotic system is extracted from the terms
corresponding to the leading order of~$\underline{R}_{ab}=0$,
expressed in the asymptotic variables~$H_{ab}$ and
coordinates~$(q,s)$, which in the harmonic case discussed in
subsection~\ref{sec:AsymptoticSystemLindRod}
is~$O(R^{-2})$. Thus one can see that
setting~$F^a\simeq\mathring{\nabla}_bF^a=O(R^{-3})$ ensures
that the gauge source functions will not contribute to the asymptotic
system. Furthermore, with such a choice of~$F^a$, the GHG asymptotic
constraint equations, i.e. the asymptotic expressions for~$C^a$,
coincide with the harmonic case since the latter are obtained from the
asymptotic expression for~$\Gamma^a$ whose leading order
is~$O(R^{-1})$. On the other hand, observe that~$W_{ab}$
and~$\mathcal{W}_{ab}$ do contribute to the asymptotic system since
their components are either quadratic in~$C_a$ or proportional
to~$R^{-1}C_a$.

Computing the asymptotic system for
equation~\eqref{eq:ReducedEFEGasHil} one obtains
\begin{flalign} \label{eq:AsymptSystemGasHil}
  (2 \mathring{\nabla}_{s} - H_{LL}{}
  \mathring{\nabla}_{q})\mathring{\nabla}_{q}H_{ab} = T_{ab},
\end{flalign}
where~$T_{ab}$ is expressed in the flat null frame
as~\eqref{eq:SymmetricTensorInFlatNullFrame} where the only
non-vanishing components are
\begin{align}
T_{LL}{} &= \big(-2\omega + \partial_{q}H_{LL}{}\big)
\partial_{q}H_{LL}{}, \nonumber\\  T_{LS_ A}{} &= \; 
\big(-2\omega +  \partial_{q}H_{LL}{}\big)
\partial_{q}H_{LS_ A}{}, \nonumber  \\
T_{S_{ A} S_{ B}}{} &= \;\delta_{ A B}
\big(-\omega + \tfrac{1}{2} \partial_{q}H_{LL}{}\big)
(\partial_{q}H^{\varnothing}{}) \nonumber\\
T_{\underline{L}\underline{L}}
&= \; - 4 (\partial_{q}H^{+}{})^2-4 (\partial_{q}H^{\times})^2 
\end{align}
 On the other
hand, as explained above, the asymptotic GHG constraint condition is
identical to equation~\eqref{eq:AsymptHarmonicCondition} for our
choice of~$F^a$. A first observation is that contracting
with~$\underline{L}^a\underline{L}^b$
equation~\eqref{eq:SimplifiedEquationForBadComponent} is of course
recovered.  The term~$W_{ab}$ was included precisely for this
purpose. The advantage of this approach is that the above succinct
expression for~$T_{\underline{L}\underline{L}}$ is obtained directly
from the evolution equations without assuming that~$C^a=0$. As it will
be discussed in greater detail in the next subsection, the
parameter~$\omega$ in the term~$\mathcal{W}_{ab}$ was included so that
the evolution equations ensure the small constraint violations are
damped away along the integral curves of then vector field~$2
\partial_{s} -H_{LL}{} \partial_{q}$ while the terms quadratic
in~$C_a$ in with $\mathcal{W}_{ab}$ are introduced to be able to
commute~$2 \partial_{s} -H_{LL}{} \partial_{q}$ and~$\partial_{q}$
when necessary. A detailed discussion on how to exploit these
constraint additions in view of a future numerical implementation
using the dual foliation formalism is the content of the next
subsection.

%%%%%%%%%%%%%%%%%%%%%%%%%%%%%%%%%%%%%%%%%%%%%%%%%%%%%%%%%%%%%%%%%%%%%%%%%%%%%%%%%%%%%%%
\subsection{Analysis of the asymptotic system}\label{sec:AnaAsympSys}
%%%%%%%%%%%%%%%%%%%%%%%%%%%%%%%%%%%%%%%%%%%%%%%%%%%%%%%%%%%%%%%%%%%%%%%%%%%%%%%%%%%%%%%

A direct computation then shows that the asymptotic
system~\eqref{eq:AsymptSystemGasHil} can be expressed as
\begin{subequations}\label{eq:goodbadugly}
\begin{align}
& ( \partial_{s} - \tfrac{H_{LL}}{2}
\partial_{q})\partial_{q}H_{\mathcal{U}}{}  =  -\omega
\partial_{q}H_{\mathcal{U}}{} {} + {}
\tfrac{1}{2}\partial_{q}H_{\mathcal{U}}{}\partial_{q}H_{LL}, \label{eq:thegood}\\
& (\partial_{s} - \tfrac{H_{LL}}{2}
\partial_{q})\partial_{q}H_{\mathcal{V}}{}  = 
0 , \label{eq:theugly}\\
& ( \partial_{s} - \tfrac{H_{LL}}{2}
\partial_{q})\partial_{q}H_{\underline{L}\underline{L}}  = 
-2 (\partial_{q}H^{+})^2 - 2(\partial_{q}H^{\times})^2,
\label{eq:thebad} 
\end{align}
\end{subequations}
where~$\mathcal{V} \in\{ \ L\underline{L}, \underline{L}S_ A,
\times, + \}$ and $\mathcal{U} \in \{LL, LS_ A,\varnothing \}$.

Observe the asymptotic equations for the components of~$H_{ab}$, split
in three classes whereas in the asymptotic system of
subsection~\ref{sec:AsymptoticSystemLindRod} they split only in two. The
components of~$H_{ab}$ satisfying equation~\eqref{eq:theugly} will be
referred as the ``the good'' components, on the other hand,
$H_{\underline{L}\underline{L}}$ will be referred as ``the bad''
component and finally, those satisfying equation~\eqref{eq:thegood} as
the ``the ugly'' components. Observe that the asymptotic
system~\eqref{eq:goodbadugly} does not follow the hierarchical
structure introduced in~\cite{Kei17}. In particular, the equation
for~$H_{LL}$ does not lie on the bottom level of the hierarchy and its
equation contains quadratic non-linearities of~$H_{LL}$ itself.
Nevertheless we will see in what follows that one can still integrate
these equations and obtain, under suitable assumptions on the initial
data, bounded solutions for~$H_{\mathcal{U}}$ and moreover,
that~$\partial_qH_{\mathcal{U}} \sim O(R^{-\omega})$.
  
For completeness, observe that, with the current decay assumption on
the gauge source functions~$F^a$, the asymptotic GHG constraint
equations are identical to those given in
equations~\eqref{eq:AsymptHarmonicConditionComps} and can be compactly
written as
\begin{align}\label{eq:AsymptConstraints}
C_{\mathcal{U}}=0,
\end{align}
where~$C_{\mathcal{U}}\equiv
\partial_{q}H_{\mathcal{U}}$. Nevertheless, recall, that the
conditions encoded in~\eqref{eq:AsymptConstraints} are not expected to
be fulfilled in a free evolution numerical implementation as one
expects small violations to the GHG constraints.  Consequently, the
following analysis will be done without assuming that
equation~\eqref{eq:AsymptConstraints} is satisfied.

%%%%%%%%%%%%%%%%%%%%%%%%%%%%%%%%%%%%%%%%%%%%%%%%%%%%%%%%%%%%%%%%%%%%%%%%%%%%%%%%%%%%%%%
\paragraph*{Solution for the {\it uglies}:} Observe that the equations
encoded in~\eqref{eq:thegood} are the asymptotic equations for the
components of the metric perturbation associated with the GHG
constraints. In other words, equation~\eqref{eq:thegood} can be read
as the following equation for the GHG constraints
\begin{align}\label{eq:goodeq-derivative}
( \partial_{s} - \frac{1}{2}H_{LL}{}
\partial_{q})C_{\mathcal{U}} = -\omega 
C_{\mathcal{U}} + \frac{1}{2} 
C_{\mathcal{U}}{} C_{LL}.
\end{align}
Let~$ \gamma=(s(\lambda),q(\lambda),\theta^A(\lambda))$ be a curve
passing through the
point~$ \gamma_\star=(s_\star,q_\star,\theta^A_\star)$,
where~$s_\star=s(\lambda_\star)$, $q_\star=q(\lambda_\star)$
and~$\theta^A_\star=\theta^A(\lambda_\star)$, satisfying
\begin{align}\label{eq:eqCharacteristicCurve}
\frac{\mbox{d}s}{\mbox{d}\lambda}=1,\qquad
\frac{\mbox{d}q}{\mbox{d}\lambda}=-\frac{1}{2}H_{LL}{},
\end{align}
and~$\theta^A(\lambda)=\theta^A_\star$. Observe that integrating the
first equation in~\eqref{eq:eqCharacteristicCurve} and
setting~$\lambda_\star=s_{\star}$ one has~$s=\lambda$. Thus, one can
use~$s$ to parameterize the curve
as~$ \gamma=(s,q(s),\theta^A_\star)$. Then, considering
equation~\eqref{eq:goodeq-derivative} along~$ \gamma$ one has
\begin{align}\label{eq:goodeq-derivative-along-char}
\frac{\mbox{d}C_{\mathcal{U}}( \gamma)}{\mbox{d}s} =
-\omega  C_{\mathcal{U}}( \gamma) +
\frac{1}{2} C_{\mathcal{U}}( \gamma) C_{LL}( \gamma).
\end{align}
Solving equation~\eqref{eq:goodeq-derivative-along-char} for
${\mathcal{U}}={LL}$ one obtains,
\begin{align}
C_{LL}( \gamma)=\frac{2\omega}{1-e^{\omega(p+2A_{\star})}},
\end{align}
where~$A_{\star}$ is an integration constant. Using that~$\omega > 0$
and denoting the initial datum~$C_{LL}(s_\star,q_\star)$
as~$C_{LL}( \gamma_\star)$ then the solution can be written as
\begin{align}\label{eq:CL-sol}
C_{LL}( \gamma)=\frac{2 C_{LL}( \gamma_\star)\omega
e^{\omega s_{\star}}}{C_{LL}( \gamma_\star)( e^{\omega
s_{\star}} - e^{\omega s} ) + 2\omega e^{\omega s}}.
\end{align}
This equation is dangerous since one could have blow up at finite
value of~$s$. Nevertheless, assuming that~$C_{LL}(
\gamma_\star)<2\omega$ the denominator in expression~\eqref{eq:CL-sol}
is positive, and then~$C_{LL}( \gamma)$ is bounded. Moreover, one has
that
\begin{align}
\lim_{s\rightarrow \infty} C_{LL}( \gamma)=0.\nonumber
\end{align}
Now, we can use expression \eqref{eq:CL-sol} to solve
equation~\eqref{eq:goodeq-derivative} for $C_{\mathcal{U}}$
with~$\mathcal{U} \in \{LS_ A, \varnothing \}$. A direct calculation
yields,
\begin{align}\label{eq:CU-sol}
C_{\mathcal{U}}( \gamma)=\frac{2\mathcal{C}_{\mathcal{U}}( \gamma_\star)\omega
e^{\omega s_{\star}}}{\mathcal{C}_{LL}( \gamma_\star)(e^{\omega s_{\star}}-e^{\omega s})
+2\omega e^{\omega s} }.
\end{align}  
Again observe that
\begin{align}
\lim_{s\rightarrow \infty} C_{\mathcal{U}}( \gamma)=0.
\end{align}
Consequently, recalling that~$C_{\mathcal{U}} \equiv
\partial_qH_{\mathcal{U}}$, then, one concludes
that,~$\partial_qH_{\mathcal{U}}( \gamma)$ is determined by
equations~\eqref{eq:CL-sol} and~\eqref{eq:CU-sol}. Now, to solve
for~$H_{\mathcal{U}}$ observe that
\begin{multline}
\partial_q(2\partial_{s}-H_{LL}\partial_{q})H_{\mathcal{U}}=
(2\partial_{s}-H_{LL}\partial_{q})\partial_qH_{\mathcal{U}}
\\ -\partial_{q}H_{LL}\partial_qH_{\mathcal{U}}.
\end{multline}
Using the latter identity with equation~\eqref{eq:thegood} we get
\begin{align}
  \partial_q\big[(2\partial_{s}-H_{LL}\partial_{q})H_{\mathcal{U}}\big]=
  -2\omega\partial_{q}H_{\mathcal{U}}.
\end{align}
Consequently, integrating from~$q_0$ to~$q$ we have
\begin{align}\label{eq:GoodH-FirstIntegral}
(2\partial_{s}-H_{LL}\partial_{q})H_{\mathcal{U}}+2\omega
H_{\mathcal{U}} = \mathcal{G}_{\mathcal{U}}(s),
\end{align}
where
\begin{align}
  \mathcal{G}_{\mathcal{U}}(s) \equiv
  ((2\partial_{s}-H_{LL}\partial_{q})H_{\mathcal{U}}+2\omega
  H_{\mathcal{U}})|_{\mathcal{S}},
\end{align}
and~$\mathcal{S}$ is the hypersurface determined by $T=0$. Observe
that if~$p \in \mathcal{S}$, then $p$ has coordinates~$(s,q_0(s),
\theta^A)$ where~$q_0(s)=e^s$.

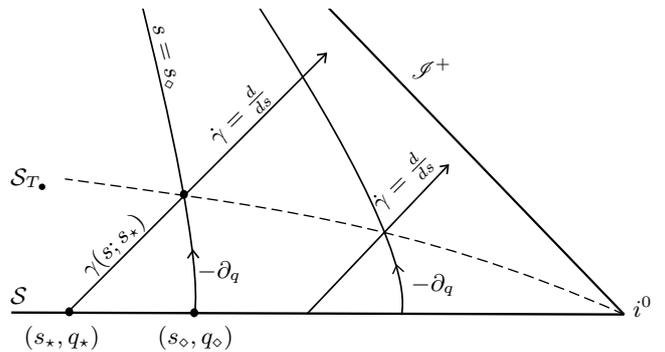
\begin{figure}
\includegraphics[width=0.45\textwidth]{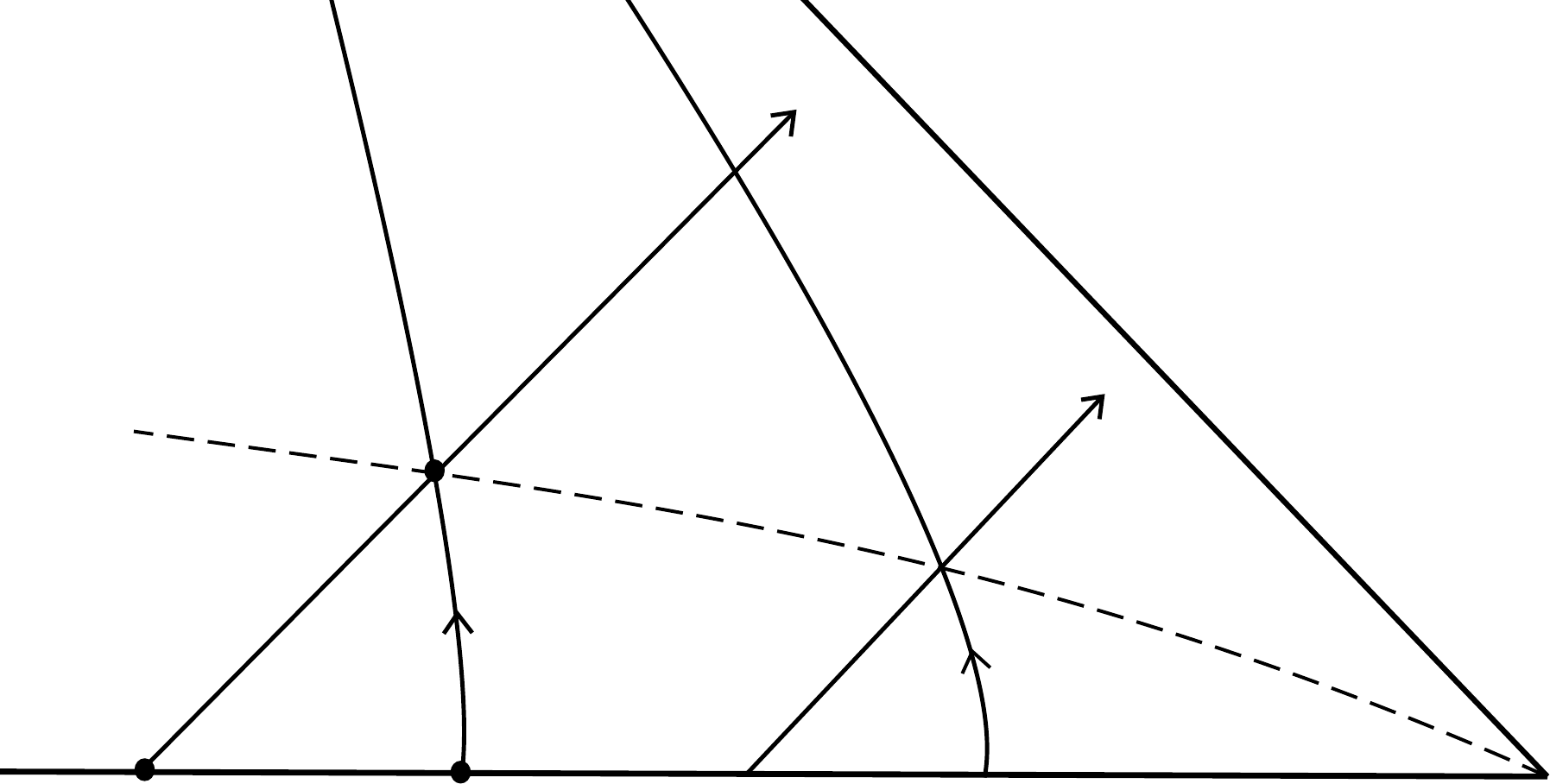}
\put(-225,-10){$(s_\star,q_\star)$} \put(-175,-10){$(s_\diamond,q_\diamond)$}
\put(-160,15){$-\partial_q$}
\put(-80,10){$-\partial_q$}
\put(-158,65){\rotatebox{45}{$\dot{ \gamma}=\tfrac{d}{ds}$}}
\put(-97,40){\rotatebox{45}{$\dot{ \gamma}=\tfrac{d}{ds}$}}
\put(-205, 15){\rotatebox{45}{$ \gamma(s;s_\star)$}}
\put(-178, 110){\rotatebox{-75}{$s=s_\diamond$}}
\put(-230, 5){$\mathcal{S}$}
\put(-230, 50){$\mathcal{S}_{T_\bullet}$}
\put(-80, 90){$\mathscr{I}^+$}
\put(3, 0){${i}^0$}
\caption{Schematic depiction of the integration of the components
  of~$H_{ab}$ associated with the constraints. The angular
  coordinates~$\theta^A_{\star}$ have been suppressed to simplify the
  notation in this figure and this caption. Given a
  point~$(s_\star,q_\star)$ on the initial hypersurface~$\mathcal{S}$,
  the characteristic curve that passes through that point is denoted
  as~$ \gamma(s;s_{\star})$. Observe that varying~$s_\star$, then
  one obtains a congruence of curves parameterized by $s_\star$. The
  diagram shows the hypersurface~$S_{T_{\bullet}}$ determined by the
  condition~$T=T_\bullet$ where, in turn, $T_{\bullet}$, is determined
  by the intersection of the curve with constant~$s=s_{\diamond}$ and
  the characteristic~$ \gamma(s;s_{\star})$ as shown in the
  diagram. The integral curves of~$\partial_q$ are the curves of
  constant~$s$ along which the first integration is performed. Then,
  the second integration is performed along the characteristic
  curves~$ \gamma(s;s_{\star})$} \label{fig:DiagramIntegration}
\end{figure}

Now, consider the curve~$ \gamma$ intersecting the
hypersurface~$\mathcal{S}$ at the point~$ p_\star$ with
coordinates~$(s_\star,q_{\star},\theta^A_\star)$,
where~$q_\star=e^{s_\star}$. Proceeding similarly as in
equation~\eqref{eq:goodeq-derivative-along-char} one can rewrite
equation~\eqref{eq:GoodH-FirstIntegral} as
\begin{align}
 \frac{\mbox{d}H_{\mathcal{U}}( \gamma)}{\mbox{d}s} + \omega
 H_{\mathcal{U}}( \gamma)=\frac{1}{2} \mathcal{G}_{\mathcal{U}}(s).
\end{align}  
The latter equation can be solved to yield
\begin{align}\label{eq:HU-sol}
H_{\mathcal{U}}( \gamma)=e^{-\omega s} \bigg(e^{\omega s_\star}
  H_{\mathcal{U}}( \gamma_\star)-\frac{1}{2}\int_{s_\star}^{s}e^{\omega
    \bar{s}}\mathcal{G}_{\mathcal{U}}(\bar{s})\mbox{d}\bar{s}\bigg).
\end{align}
Thus, $H_{\mathcal{U}}( \gamma)$
and~$\partial_{q}H_{\mathcal{U}}( \gamma)$ are determined by data on
$\mathcal{S}$ via equations \eqref{eq:HU-sol}, \eqref{eq:CL-sol} and
\eqref{eq:CU-sol}. Additionally, observe that, from
equation~\eqref{eq:GoodH-FirstIntegral} one has
\begin{align}
\partial_{s}H_{\mathcal{U}}=\frac{1}{2}H_{LL}( \gamma)\partial_q
H_{\mathcal{U}}( \gamma) -\omega H_{\mathcal{U}} + \frac{1}{2}
\mathcal{G}_{\mathcal{U}}(s).
\end{align}
Therefore, $\partial_{s}H_{\mathcal{U}}( \gamma)$ is determined
by~$H_{\mathcal{U}}( \gamma)$,~$\partial_{q}H_{\mathcal{U}}( \gamma)$
and~$\mathcal{G}_{\mathcal{U}}(s)$. Finally, recalling that~$s=\ln R$,
then, if the initial data on~$\mathcal{S}$ is such that
\begin{align}
H_{\mathcal{U}}|_{\mathcal{S}} & \sim O(1), &
&\text{sup}( \partial_{q}H_{LL}|_{\mathcal{S}}) < 2\omega. \nonumber   \\ 
\partial_{q}H_{\mathcal{U}}|_{\mathcal{S}} &\sim O(R^{-1}), &
&\partial_{s}H_{\mathcal{U}}|_{\mathcal{S}} \sim O(R^{-1}), 
\end{align}
Then $\mathcal{G}_{\mathcal{U}}(s)\simeq O(1)$ and, consequently,
using equations~\eqref{eq:HU-sol},\eqref{eq:CL-sol}
and~\eqref{eq:CU-sol} one concludes that
\begin{flalign}
\partial_{q}H_{\mathcal{U}}( \gamma) &\sim O(R^{-\omega}), \quad
\partial_{s}H_{\mathcal{U}}( \gamma) \sim O(1), \nonumber \\
H_{\mathcal{U}}( \gamma) &\sim O(1).
\end{flalign}
Observe that, since $\partial_R=\frac{1}{R}\partial_s$ then
$\partial_{s}H_{\mathcal{U}}( \gamma) \sim O(1)$ implies that
$\partial_{R}H_{\mathcal{U}}( \gamma) \sim O(R^{-1})$. The latter
decay rates imply that the coordinate light-speed condition, as
expressed in equation \eqref{eq:DFcondition}, is satisfied if one sets
$\omega\geq \delta$.

%%%%%%%%%%%%%%%%%%%%%%%%%%%%%%%%%%%%%%%%%%%%%%%%%%%%%%%%%%%%%%%%%%%%%%%%%%%%%%%%%%%%%%%
\paragraph*{Solution for the {\it goods}:} Notice that in
contrast with the case of equation~\eqref{eq:thegood}, for
equation~\eqref{eq:theugly} the fields~$\partial_{q}H_{\mathcal{V}}$
are not associated to the constraints. Nevertheless, along the curve~$
\gamma$ equation~\eqref{eq:theugly} simply reads
\begin{align}
\frac{\mbox{d}}{\mbox{d}s}\partial_{q}H_{\mathcal{V}}( \gamma)=0.
\end{align}
Thus
\begin{align}\label{eq:FirstIntegralGood}
\partial_{q}H_{\mathcal{V}}( \gamma)=\partial_{q}H_{\mathcal{V}}( \gamma_\star).
\end{align}
Now, integrating the second equation in
\eqref{eq:eqCharacteristicCurve} one has that
\begin{align}
q=q_\star -\frac{1}{2}\int_{s_\star}^sH_{LL}(\bar{s})\mbox{d}\bar{s}.
\end{align}
Observe
that~$\partial_{q}H_{\mathcal{V}}( \gamma_\star)=\partial_{q}H_{\mathcal{V}}(s_\star,
q_\star)$ does not depend on~$s$.  Using~$q_\star$ as a new
coordinate. 
Namely, considering a coordinate system $(s,u)$ where
$u=q+\frac{1}{2}\int_{s_\star}^sH_{LL}(\bar{s})\mbox{d}\bar{s}$ then
$\partial_{q}H_{\mathcal{V}}( \gamma_\star)$ is only a function of
$u$.  Moreover, since
$\partial_{q}H_{\mathcal{V}}=\partial_{u}H_{\mathcal{V}}$ then,
integrating in $u$ one has
\begin{align}
H_{\mathcal{V}}( \gamma)=\int_{u_\star}^{u}\partial_{q}H_{\mathcal{V}}(\bar{u})\mbox{d}\bar{u}.
\end{align}
Thus, $H_{\mathcal{V}}( \gamma)$ does not depend on~$s$,
consequently,~$H_{\mathcal{V}}( \gamma) \sim O(1)$.

%%%%%%%%%%%%%%%%%%%%%%%%%%%%%%%%%%%%%%%%%%%%%%%%%%%%%%%%%%%%%%%%%%%%%%%%%%%%%%%%%%%%%%%
\paragraph*{Solution for the {\it bad}:}To integrate
equation~\eqref{eq:thebad} observe that along the curve~$ \gamma$
one has
\begin{align}
\frac{\mbox{d}}{\mbox{d}s}\partial_{q}H_{\underline{L}\underline{L}}( \gamma)=
-2\big[(\partial_{q}H^{+})^2-2(\partial_{q}H^{\times})^2\big].
\end{align}
Using the previous results obtained for~$H_\mathcal{V}$, one has
that~$\partial_{q}H^{+}$ and~$\partial_{q}H^{\times}$ do not depend
on~$s$. Thus, integrating in~$s$, one obtains
\begin{align}
\partial_{q}H_{\underline{L}\underline{L}}( \gamma)=
-2\big[(\partial_{q}H^{+}( \gamma_\star))^2
+(\partial_{q}H^{\times}( \gamma_\star))^2\big]s.
\end{align}
Then, using the coordinates~$(s,u)$ and that~$\partial_q=\partial_u$,
one has that
\begin{align}
H_{\underline{L}\underline{L}}( \gamma)=
-2s\int_{u_\star}^{u}((\partial_{q}H^{+}( \gamma_\star))^2
+(\partial_{q}H^{\times}( \gamma_\star))^2)\mbox{d}\bar{u}.
\end{align}
Observe that the integration of the asymptotic equation
for the goods and the bad is analogous to that of the
model equation \eqref{ToyModelIII}.

%%%%%%%%%%%%%%%%%%%%%%%%%%%%%%%%%%%%%%%%%%%%%%%%%%%%%%%%%%%%%%%%%%%%%%%%%%%%%%%%%%%%%%%
\subsection{The eikonal equation}\label{eq:eikonalequationAsymptotic}
%%%%%%%%%%%%%%%%%%%%%%%%%%%%%%%%%%%%%%%%%%%%%%%%%%%%%%%%%%%%%%%%%%%%%%%%%%%%%%%%%%%%%%%

In this subsection the asymptotic form of the eikonal equation and its
relation to the characteristics of the asymptotic
system~\eqref{eq:AsymptSystemGasHil} are discussed.
As emphasized before, when dealing with asymptotic computations
we will use the Minkowski metric $m_{ab}$ to raise and lower
indices in all tensors except for $g_{ab}$ in which case we use equation
\eqref{eq:upsplitg}.

%%%%%%%%%%%%%%%%%%%%%%%%%%%%%%%%%%%%%%%%%%%%%%%%%%%%%%%%%%%%%%%%%%%%%%%%%%%%%%%%%%%%%%%
\paragraph*{The eikonal equation and its asymptotic system:}
The vectors $L^a$ and $\underline{L}^a$ are null vectors in Minkowski
spacetime,
namely~$m^{ab}L_{a}L_{b}=m^{ab}\underline{L}_{a}\underline{L}_{b}=0$.
In what follows, the asymptotic form of vectors~$\xi^a$ which are null
respect to the perturbed metric~$g_{ab}$ will be discussed.

We start by recalling that the eikonal equation reads
\begin{align}\label{eq:eikonal}
g^{ab}\nabla_{a}u\nabla_{b}u=0,
\end{align}
and that if~$\xi^a=g^{ab}\nabla_{b}u$, then eikonal equation
simply states that~$\xi^a$ is null respect to~$g_{ab}$. Now, let
\begin{align}\label{eq:split-true-null-vector}
  \xi^a=\mathring{\xi}^a+\frac{1}{R}\check{\xi}^a,
\end{align}
where
\begin{align}\label{eq:flat-null-vector-cond}
m_{ab}\mathring{\xi}^a\mathring{\xi}^b=0.
\end{align}
Using this notation one has that
\begin{align}
g_{ab}\xi^a\xi^b &= \frac{H_{ab}  \mathring{\xi}^{a}
\mathring{\xi}^{b}}{R} + \frac{ 2 m_{ab} \mathring{\xi}^{a}
\check{\xi}^{b}}{R} + \ \frac{ 2H_{ab} \mathring{\xi}^{a}
  \check{\xi}^{b}}{R^2} \nonumber\\
&+ \frac{ m_{ab} \check{\xi}^{a} \check{\xi}^{b}}{R^2} +
\frac{\ H_{ab} \check{\xi}^{a} \ \check{\xi}^{b}}{R^3}.
\end{align}
Thus, $\xi^a$ is a null vector respect to~$g_{ab}$ to leading
order~$R^{-1}$, if the following condition is satisfied
\begin{align}\label{eq:nullconditioncomponent}
H_{ab} \mathring{\xi}^{a} \mathring{\xi}^{b} + \ 2 m_{ab}
\mathring{\xi}^{a} \check{\xi}^{b} = 0.
\end{align}
Naturally, one can algebraically impose that~$g_{ab}\xi^a\xi^b=0$ also
to order~$R^{-2}$ and~$R^{-3}$. Nevertheless, consistent with the
discussion of the asymptotic system, one is interested only in the
leading order. In the following, it will be shown that, if
equation~\eqref{eq:nullconditioncomponent} is satisfied on the initial
hypersurface~$\mathcal{S}$ then~$\xi^a$ will be, to leading order,
null respect to~$g_{ab}$ in~$\mathcal{D}(\mathcal{S})\subseteq
\mathcal{M}$ provided that the coordinate light-speed condition of
subsection~\ref{sec:coordlight-speed} is satisfied. The condition imposed
by the eikonal equation~\eqref{eq:eikonal} can be formulated as an
initial value problem as follows
\begin{align}\label{eq:eikonal-initial-value-problem}
\nabla_{c}(g_{ab}\xi^a\xi^b)=0, \qquad (g_{ab}\xi^a\xi^b)|_{\mathcal{S}}=0.
\end{align}

The initial condition~$(g_{ab}\xi^a\xi^b)|_{\mathcal{S}}=0$ to leading
order~$R^{-1}$ corresponds to
equation~\eqref{eq:nullconditioncomponent} where the fields are
restricted to the initial hypersurface~$\mathcal{S}$.

Now observe that the condition~$\xi^a=g^{ab}\nabla_bu$ is equivalent
to the integrability condition~$\nabla_{[a}\xi_{b]}=0$. Using this
integrability condition, the first equation
in~\eqref{eq:eikonal-initial-value-problem} reduces to the geodesic
equation
\begin{align}\label{eq:geodesic-true}
  \xi^b\nabla_b\xi^a=0.
\end{align}
Using the integrability
condition~$\mathring{\nabla}_{[a}\mathring{\xi}_{b]}=0$ and proceeding
in analogous way with equation~\eqref{eq:flat-null-vector-cond} one
has,
\begin{align}\label{eq:geodesic-flat}
  \mathring{\xi}^b\mathring{\nabla}_b\mathring{\xi}^a=0.
\end{align}
Notice that the geodesic equation~\eqref{eq:geodesic-true} can be
expressed more explicitly, using the covariant
derivative~$\mathring{\nabla}$, as
\begin{align}
  2\xi^a \mathring{\nabla}_a\xi^c
  + \xi^a\xi^b(g)^{cd} (\mathring{\nabla}_ag_{bd} +
  \mathring{\nabla}_bg_{ad} - \mathring{\nabla}_dg_{ab}) = 0.
  \label{eq:geodesic-explicit}
\end{align}
Using equations~\eqref{eq:splitg}, \eqref{eq:AsymptSystemRecipe},
\eqref{eq:split-true-null-vector}, \eqref{eq:geodesic-true}
and~\eqref{eq:geodesic-flat} one obtains
\begin{multline}\label{eq:asympt-system-geodesic-expansion}
\xi^a\nabla_a\xi^c \simeq \frac{1}{R}\big( 2 \mathring{\xi}^{a}
\ \mathring{\nabla}_{a}\check{\xi}^{c}
+ 2 \check{\xi}^{a} \ \mathring{\nabla}_{a}\mathring{\xi}^{c} +
m^{cd} ( \mathring{\xi}^{a} \mathring{\xi}^{b} \ \mathring{\nabla}_{a}H_{bd}  \\
+ \mathring{\xi}^{a} \ \mathring{\xi}^{b} \mathring{\nabla}_{b}H_{ad} -
\mathring{\xi}^{a} \mathring{\xi}^{b} \ \mathring{\nabla}_{d}H_{ab}) \big).
\end{multline}
Thus, similarly as it was done for the Einstein field equations, one
says that the asymptotic system for equation~\eqref{eq:geodesic-true}
is encoded in the leading order of the above expansion. Recall
that~$\mathring{\xi}^a$ represents any null vector with respect to
$m_{ab}$. Choosing~$\mathring{\xi}^a=L^a$ and contracting with the
flat null frame one obtains that
\begin{align}
\underline{\hat{L}}_c\xi^a\nabla_a\xi^c & \simeq \frac{2}{R}
\mathring{\nabla}_qH_{LL}, \nonumber \\
\hat{L}_c\xi^a\nabla_a\xi^c &= O(R^{-2}), \nonumber \\
{\hat{S}{}^{ A}}_c\xi^a\nabla_a\xi^c &= O(R^{-2}).\label{eq:asympt-geodesic}
\end{align}
Recalling from subsection~\ref{sec:AnaAsympSys} that, along the
characteristic curve~$ \gamma$, one
has~$\mathring{\nabla}_qH_{\mathcal{U}} \simeq O(R^{-\omega})$ we
observe that setting~$\omega\geq 1$ ensures that
equation~\eqref{eq:geodesic-explicit} is satisfied to leading
order~$R^{-1}$.

Therefore, exploiting the discussion of subsection~\ref{sec:AnaAsympSys}
and setting~$\omega \geq 1$, one can conclude that
\begin{align}\label{eq:eikonalInitialData}
  (g_{ab}\xi^a\xi^b)|_\mathcal{S}=0,
\end{align}
implies that, to leading order,
\begin{align}
  g_{ab}\xi^a\xi^b=0, \qquad \text{in} \qquad
  \mathcal{D}(\mathcal{S})\subseteq \mathcal{M}.
\end{align}

Additionally, observe that for~$\mathring{\xi}^a=L^a$, using
equation~\eqref{eq:nullconditioncomponent},
condition~\eqref{eq:eikonalInitialData} reduces to
\begin{align}
(\hat{L}_a\check{\xi}^a)|_{\mathcal{S}}=-\frac{1}{2}H_{LL}|_{\mathcal{S}}.
\end{align}
By the argument above, to leading order, one has~$g_{ab}\xi^a\xi^b=0$
in~$\mathcal{D}(\mathcal{S})\subseteq \mathcal{M}$. Using
equation~\eqref{eq:nullconditioncomponent}, this is equivalent to
\begin{align}
  \hat{L}_a\check{\xi}^a=-\frac{1}{2}H_{LL}, \qquad \text{in}
  \qquad \mathcal{D}(\mathcal{S})\subseteq \mathcal{M}.
\end{align}
 
%%%%%%%%%%%%%%%%%%%%%%%%%%%%%%%%%%%%%%%%%%%%%%%%%%%%%%%%%%%%%%%%%%%%%%%%%%%%%%%%%%%%%%%
\paragraph*{The characteristics of the asymptotic system and null
  geodesics:} We now show that the characteristics of the asymptotic
system~\eqref{eq:AsymptSystemGasHil} correspond, to leading order, to
null geodesics of the perturbed
spacetime~$(g_{ab},\mathcal{M})$. Observe that, using
equations~\eqref{eq:NullFrameToSpFrame}
and~\eqref{eq:AsymptSystemRecipe} one has that
\begin{align}
(\partial_q)^a=-\frac{1}{2}(L^a+\underline{L}^a), \qquad
(\partial_s)^a = R L^a.
\end{align}
We define 
\begin{align}
\chi^a = \frac{1}{R}\big((\partial_s)^a
-\frac{1}{2}H_{LL}(\partial_q)^a\big),
\end{align}
and observe that $\chi^a$ is a tangent to the curve $ \gamma$.
Furthermore, notice that $\chi^a$ can be expressed as
\begin{align}
\chi^a = L^a + \frac{1}{4R}H_{LL}(L^a+\underline{L}^a).
\end{align}
Now, consistent with equation~\eqref{eq:split-true-null-vector},
consider the
split~$\chi^a=\mathring{\chi}^a+\frac{1}{R}\check{\chi}^a$
with~$\mathring{\chi}^a=L^a$ and
\begin{align}
\check{\chi}^a= \frac{1}{4}H_{LL}(L^a+\underline{L}^a).
\end{align}
Then, one readily observes that
\begin{align}
\hat{L}_a\check{\chi}^a=-\frac{1}{2}H_{LL}.
\end{align}
Consequently, using the results of
subsection~\ref{eq:eikonalequationAsymptotic}, one concludes that the
vector~$\chi^a$ representing the tangent vector to the characteristics
of the asymptotic system~\eqref{eq:AsymptSystemGasHil}, is, to leading
order, a null vector in the perturbed spacetime~$(g_{ab},\mathcal{M})$.
 
%%%%%%%%%%%%%%%%%%%%%%%%%%%%%%%%%%%%%%%%%%%%%%%%%%%%%%%%%%%%%%%%%%%%%%%%%%%%%%%%%%%%%%%
\section{The asymptotic system for first order GHG with constraint
  damping}\label{sec:AsympSysFirstOrderGHG}
%%%%%%%%%%%%%%%%%%%%%%%%%%%%%%%%%%%%%%%%%%%%%%%%%%%%%%%%%%%%%%%%%%%%%%%%%%%%%%%%%%%%%%%

In this section, the first order GHG Einstein evolution equations with
the constraint additions introduced in
section~\ref{sec:GasHilConstraintAddition} are presented and its
asymptotic system is derived in analogous way as that of the model
equation~\eqref{ToyModelIII}. In subsection~\ref{EFEGHGnonasympt} the
general evolution equations are given, and, since this is not an
asymptotic computation, the usual conventions for raising and lowering
indices using~$g_{ab}$ is employed. In contrast, in
subsection~\ref{EFEGHGasympt} the corresponding asymptotic system is
derived, and, consequently, the indices are handled in concordance
with the asymptotic calculations of
section~\ref{sec:AsymptoticSystemLindRod}.

%%%%%%%%%%%%%%%%%%%%%%%%%%%%%%%%%%%%%%%%%%%%%%%%%%%%%%%%%%%%%%%%%%%%%%%%%%%%%%%%%%%%%%%
\subsection{First order GHG}\label{EFEGHGnonasympt}
%%%%%%%%%%%%%%%%%%%%%%%%%%%%%%%%%%%%%%%%%%%%%%%%%%%%%%%%%%%%%%%%%%%%%%%%%%%%%%%%%%%%%%%

Consider the reduced Ricci operator as given in
equation~\eqref{ReducedRicciTensorDefinitionGeneral}, where
$\mathcal{T}_{ab}$ denotes a generic constraint addition. Then, using
the foregoing conventions by expressing the derivatives in terms of
$\mathring{\nabla}$, the GHG evolution equations~\cite{HilHarBug16},
in abstract index notation, read
\begin{flalign}
  & N^c \mathring{\nabla}_c g_{ab} =  S^{(g)}_{ab} \nonumber \\
  & N^f  \gamma_{a}{}^{d} \mathring{\nabla}_{f}\phi_{dbc} =
  {}  - 
  \gamma_{a}{}^{d} \mathring{\nabla}_{d}\pi_{bc} {}
  + \gamma_{2}{}
 \gamma_{a}{}^{d} \mathring{\nabla}_{d}g_{bc}
 + S^{(\phi)}_{abc} \nonumber \\
 &  N^c  \mathring{\nabla}_{c}\pi_{ab} =   \gamma^{cd}
 \mathring{\nabla}_{c}\phi_{dab} + S^{(\pi)}_{ab} \label{GHG3} 
\end{flalign}
where
\begin{flalign}
 & S^{(g)}_{ab} =  - \pi_{ab} \nonumber \\ & S^{(\phi)}_{abc} =  - \gamma_{2}{}
  \phi_{abc} + \tfrac{1}{2} N^{d} N^{h} \pi_{bc} \phi_{adh} +
  \gamma^{fh} N^{d} \phi_{fbc} \phi_{adh}\nonumber \\ & S^{(\pi)}_{ab} = 
  -2\nabla_{(a}F_{b)}- {} 2 \Gamma_{ach}
  \Gamma_{bdf} g^{cd} g^{hf} {} - \tfrac{1}{2} N^{c} N^{d}
  \pi_{ab} \pi_{cd}\nonumber  \\ & - 2 g^{cd} \pi_{ca} \pi_{db} - N^{d} \pi_{df}
  \gamma^{cf} \phi_{cab} + 2 g^{cd} g^{hf} \phi_{fdb} \phi_{hca} +  2\mathcal{T}_{ab}
  \end{flalign}
and
\begin{align}
\Gamma_{abc}= N_{(c} \pi_{b)a}- \tfrac{1}{2} N_{a} \pi_{bc} +
\phi_{(c|a|b)} - \tfrac{1}{2} \phi_{abc}.
\end{align}
These evolution equations were written respect to normal derivatives
to ease the subsequent discussion. Nevertheless, observe that time
derivatives and normal derivatives are related via
\begin{align}
\mathring{\nabla}_{T} = \mathcal{A} N^a\mathring{\nabla}_{a}
+\mathcal{B}^a\mathring{\nabla}_a.
\end{align}
See~\cite{LinSchKid05} for a discussion of these equations and the
appended mathematica notebooks for a detailed derivation.

Notice that in the notation of~\cite{HilHarBug16} the formulation
parameters~$\gamma_0$, $\gamma_3$ have been set to zero and, ignoring
the potential parameters in the constraint addition
term~$\mathcal{T}_{ab}$, $\gamma_2$ is the only parameter that has
been left unspecified. The evolution system \eqref{GHG3}
is subject, as in the discussion of section
\ref{sec:GasHilConstraintAddition}, to the GHG constraints
\begin{align}\label{GHGconstraints}
C^a \equiv \Gamma^a + F^a = 0,
\end{align}
and, additionally, subject to the \emph{reduction constraints}
\begin{align}\label{ReductionConstraints}
\mathcal{C}_{abc} \equiv \gamma_a{}^{d}\nabla_dg_{bc} - \phi_{abc} = 0.
\end{align}

%%%%%%%%%%%%%%%%%%%%%%%%%%%%%%%%%%%%%%%%%%%%%%%%%%%%%%%%%%%%%%%%%%%%%%%%%%%%%%%%%%%%%%%
\subsection{Evolution equations for the perturbation}\label{EFEGHGasympt}
%%%%%%%%%%%%%%%%%%%%%%%%%%%%%%%%%%%%%%%%%%%%%%%%%%%%%%%%%%%%%%%%%%%%%%%%%%%%%%%%%%%%%%%

To derive the asymptotic system we proceed as in previous sections and
set
\begin{align}
g_{ab}=m_{ab}+h_{ab}.
\end{align}
Following the discussion of the asymptotic system in
subsection~\ref{sec:AsymptoticSystemLindRod}, once the equations for the
perturbations are obtained, as for example in
equation~\eqref{eq:eveqpertHarmonic}, to derive the asymptotic system,
the metric~$g_{ab}$ is regarded just as a symmetric tensor and the
indices are raised and lowered using~$m_{ab}$. The latter implies that
one has to be very careful with the canonical position of indices
before the expansion. In particular recall that for the inverse metric
one has
\begin{align}
g^{ab} =  m^{ab} - h^{ab}  + O^{ab}(h^2)
\end{align}
where the indices in the right hand side of the above expansion
are moved using~$m$, in other words,~$h^{ab}=m^{ac}m^{db}h_{cd}$.
To simplify the notation and avoid lengthier expressions,
let~$\doteq$ denote equality up to terms of
order~$O(h^2, h\mathring{\nabla} h, (\mathring{\nabla}h)^2)$
so that one writes,
\begin{align}
g^{ab}\doteq m^{ab}-h^{ab}.
\end{align}
The latter observations imply the following expansions for the
normal and projector
\begin{flalign}\label{splitNandGamma}
  N_a &\doteq \mathring{N}_a + \check{N}_a, \nonumber  \\
  \gamma_{ab} &\doteq  \mathring{\gamma}_{ab} +
  \check{\gamma}_{ab}, \nonumber \\
  N^a \equiv (g^{-1})^{ab} N_{b}  &\doteq \mathring{N}^{a}
   +   (\check{N}^{a} - \mathring{N}_{b}
    h^{ab}), \nonumber\\ 
    \gamma_{a}{}^{b} \equiv  (g^{-1})^{cb}\gamma_{ac} 
    &\doteq \mathring{\gamma}_{a}{}^{b} + (- h^{cb}
    \mathring{\gamma}_{ac} + \check{\gamma}_{a}{}^{b}),\nonumber \\
    \gamma^{ab} \equiv (g^{-1})^{ac}(g^{-1})^{bd}\gamma_{cd}
    &\doteq \mathring{\gamma}^{ab} + ( \check{\gamma}^{ab}- h^{ca}
    \mathring{\gamma}_{c}{}^{b} - h^{db} \mathring{\gamma}^{a}{}_{d})
\end{flalign}
where,
\begin{flalign}
  \mathring{\gamma}_{ab} \equiv & \; m_{ab} + \mathring{N}_{a}
  \mathring{N}_{b} \qquad \check{\gamma}_{ab} \equiv \; 2
  \mathring{N}_{(a} \check{N}_{b)} + h_{ab} \nonumber \\ 
  \mathring{N}_{a} \equiv & \; -\tfrac{1}{2} ( L_{a} +
  \underline{L}_{a}) \qquad \check{N}_{a} \equiv \; - \tfrac{1}{2}
  \mathring{N}_{a} \mathring{N}^{b} \mathring{N}^{c} h_{bc}
\end{flalign}
and all the indices in the right-hand side of
equation~\eqref{splitNandGamma} were moved
using~$m_{ab}$. Since~$\mathring{\nabla}_cm_{ab}=0$ then,  one has that the
background values for $\pi_{ab}$, $\phi_{abc}$ and $\Gamma_{abc}$ vanish, namely,
\begin{flalign}
\mathring{\pi}_{ab} =\mathring{\phi}_{abc} = \mathring{\Gamma}_{abc} =0.
\end{flalign}
To simplify the notation we will denote the corresponding perturbations
to the latter quantities without adding
${}\check{}{}$ to each of these fields. Thus, hereafter,
$\pi_{ab}$, $\phi_{abc}$ and $\Gamma_{abc}$ will represent the
associated perturbations.  A straightforward computation renders the following
evolution equations 
\begin{flalign}
  N^{a} \mathring{\nabla}_{a}h_{bc} & \doteq -
  {\pi}_{bc}, \nonumber \\ N^{a} \gamma_{f}{}^{b}
  \mathring{\nabla}_{a}{\phi}_{bcd} & \doteq \tfrac{1}{2} N^{a}
  N^{b} {\pi}_{cd} {\phi}_{fab} + N^{a} \gamma^{bh}
  {\phi}_{bcd} {\phi}_{fah} \nonumber \\ & +
  \gamma_{2}{} (-{\phi}_{fcd} + \gamma_{f}{}^{a}
  \mathring{\nabla}_{a}h_{cd}) - \gamma_{f}{}^{a}
  \mathring{\nabla}_{a} {\pi}_{cd}, \nonumber \\ N^{c}
  \mathring{\nabla}_{c} {\pi}_{ab} & \doteq - 2 m^{cd} m^{hf}
  {\Gamma}_{ach} {\Gamma}_{bdf} - \tfrac{1}{2} N^{c} N^{d}
  {\pi}_{ab} {\pi}_{cd} \nonumber \\ & - 2 m^{cd}
  {\pi}_{ca} {\pi}_{db} \nonumber - N^{d} {\pi}_{df}
  \gamma^{cf} {\phi}_{cab} \nonumber \\ & + 2 m^{cd} m^{hf}
   {\phi}_{fdb} {\phi}_{hca} - \gamma^{cd}
  \mathring{\nabla}_{c} {\phi}_{dab} \nonumber \\ & +2
  \mathcal{T}_{ab} -2\nabla_{(a}F_{b)},\label{eveqpertPi}
 \end{flalign}
where it is understood that~$N^a$, $\gamma^{ab}$ and~$\gamma_a{}^b$
have been substituted using equation \eqref{splitNandGamma}
while~$\mathcal{T}_{ab}$ and~$\nabla_{(a}F_{b)}$ have been expanded
out to order~$ O(h^2, h\mathring{\nabla} h,
(\mathring{\nabla} h)^2)$. Observe that, in terms of the~$T$ and~$R$
coordinates one has~$\mathring{N}_a=(\mbox{d}T)_a$. Thus, for
completeness, let~$\mathring{R}_a =(\mbox{d}R)_a$.
A direct computation shows that $\mathring{N}^a{\phi}_{abc} \doteq 0$.
The latter implies that one write
\begin{align}
\phi_{abc} \doteq \mathring{R}_a \phi_{Rbc} +
S^ A{}_a \phi_{S_ A bc},
\end{align}
where~${\phi}_{Rab} \equiv \mathring{R}^c{\phi}_{cab}$
and~${\phi}_{S_ A ab} \equiv S_{ A}{}^c{\phi}_{cab}$.
To derive the asymptotic system, following the strategy described in
section~\ref{sec:FirstOrderSysAsympt}, one defines
\begin{align}\label{defcharvariables}
\sigma^+_{ab}={\pi}_{ab}+{\phi}_{Rab}, \quad
\sigma^-_{ab}= {\pi}_{ab} - {\phi}_{Rab},
\end{align}
and expresses the evolution
equations~\eqref{eveqpertPi} in terms of the
variables
\begin{align}
\{\sigma^+_{ab}, \quad \sigma^-_{ab}, \quad {\phi}_{abc}, \quad
h_{ab} \}.\nonumber
\end{align}
Introducing the rescaled variables,
\begin{flalign}
  \Sigma^+_{ab} = & R^2\sigma^+_{ab}, \qquad\quad\;\Sigma^-_{ab} =
  R\sigma^-_{ab}, \nonumber \\ \Phi_{S_ A a b} =& R^2\phi_{S_ A ab},
  \qquad\; H_{ab}= R h_{ab}. \label{RescaledVariablesEFE}
\end{flalign}
and defining
\begin{align}
\mathbb{T}_{ab} =R^2 \mathcal{T}_{ab},
\end{align}
one obtains, assuming as in section~\ref{sec:GasHilConstraintAddition}
that the gauge source functions decay sufficiently fast~$F^a \sim
R^{-3}$, the following expansions
\begin{flalign}
  & \mathring{\nabla}_qH_{ab} + \tfrac{1}{2}\Sigma^{-}_{ab} \simeq 0,
  \nonumber \\ \nonumber \\ &
  %%%%%%%%%%%%%%%%%%%%%%
  - \tfrac{1}{2} H_{LL}{}
  \mathring{\nabla}_{q}\Sigma^{-}_{ab} 
  +  \gamma_{2}{} (H_{ab} {}
  +  \tfrac{1}{2} \Sigma^{+}_{ab} 
  -  \mathring{\nabla}_{s}H_{ab}) 
  +  \mathring{\nabla}_{s}\Sigma^{-}_{ab}
  \nonumber \\ & 
  \simeq \tfrac{1}{4} \Sigma^{-}_{aL} \Sigma^{-}_{bL} -
  \tfrac{1}{2} \hat{L}_{(a} \Sigma^{-}_{b)h} \Sigma^{-}_{L}{}^{h}
  + \tfrac{1}{8}  \hat{L}_{a} \hat{L}_{b}
  \Sigma^{-}{}^{cd} \Sigma^{-}_{cd} - 2 \mathbb{T}_{ab},
  \nonumber \\ \nonumber \\ &
  %%%%%%%%%%%%%%%%%%%%%%%%%%%%%%%%%%
  \tfrac{1}{8}( H_{\underline{L} \underline{L}}{} {}
  +{} H_{LL}{} {} - {} 2
  H_{L\underline{L}}{})
  \mathring{\nabla}_{q}\Sigma^{-}_{ab} {} {}
  + {} \gamma_{2}{} (H_{ab} 
  -{} \mathring{\nabla}_{s}H_{ab} +
  \nonumber \\ &
   \tfrac{1}{2} \Sigma^{+}_{ab}) +
  2 \mathring{\nabla}_{q}\Sigma^{+}_{ab} \simeq \Sigma^{-}_{ab} +
  \tfrac{1}{16} \Sigma^{-}_{ab} (  \Sigma^{-}_{\underline{L}\underline{L}}{}
  - 3 \Sigma^{-}_{LL}{} - 2
  \Sigma^{-}_{L\underline{L}}{} ) \nonumber \\ & - \tfrac{1}{4}
  \Sigma^{-}_{aL} \Sigma^{-}_{bL} + \tfrac{1}{2} L_{(a}
  \Sigma^{-}_{b)h}\Sigma^{-}_{L}{}^{h} - \tfrac{1}{8} L_{a} L_{b}
  \Sigma^{-}{}^{cd} \Sigma^{-}_{cd} + 2 \mathbb{T}_{ab}, \nonumber
  \\ \nonumber \\ &
  %%%%%%%%%%%%%%%%%%%%%%%%%%%%%%%%%%
  \mathring{\nabla}_q\Phi_{S_ A ab}{} =
  {}-  \tfrac{1}{2} \omega_{ A}{}^{c}
  \mathring{\nabla}_{c}\Sigma^{-}_{ab}{} - {}
  \gamma_{2}{}  (\Phi_{S_ A}{}_{ab} {}-{}
  \omega_{ A}{}^{c}  \mathring{\nabla}_{c}H_{ab}).\label{AsymptsysEFEfirstOrder}
\end{flalign}
Now, to unwrap the connection of the asymptotic system in first order
and second order form, define the rescaled reduction constraints as
follows
\begin{flalign}
  \mathbb{C}_{Rab} & =R^2\mathcal{C}_{Rab}, \qquad \mathbb{C}_{S_ A
    ab} =R^2\mathcal{C}_{S_ A ab},
\end{flalign}
Then, a direct computation using
equations~\eqref{ReductionConstraints}, \eqref{eveqpertPi},
\eqref{defcharvariables} and~\eqref{RescaledVariablesEFE}, with the
current decay assumptions about the gauge source functions~$F_a$, one
obtains
\begin{flalign}
    \Sigma^-_{ab}  &\simeq -2\mathring{\nabla}_qH_{ab} , \nonumber\\ 
    \Sigma^{+}{}_{ab} & \simeq - H_{ab} -{\mathbb{C}}_{R}{}_{ab}
    + \mathring{\nabla}_{s}H_{ab}  \nonumber \\ & \qquad \qquad \qquad
    +  \tfrac{1}{8} (H_{\underline{L} \underline{L}}{}-3H_{LL}-2H_{\underline{L}L})
  \mathring{\nabla}_{q}H_{ab},
  \nonumber\\ \Phi_{S_ A ab} & =
  -\mathbb{C}_{S_ A}+\omega_{ A}{}^c\mathring{\nabla}_cH_{ab},
  \label{RescaledVarsInTermsOfReductionConstraints}
\end{flalign}
where, written in full, these quantities satisfy,
\begin{align}\label{RelationSigmaPlusWithTheRadialReductionConstraint}
\Sigma^-_{ab}+\frac{\Sigma^+_{ab}}{R}=2\mathring{\nabla}_qH_{ab}
+\frac{2\mathring{\nabla}_sH_{ab}}{R} -
\frac{2H_{ab}}{R}-\frac{2\mathbb{C}_{Rab}}{R}.
\end{align}
Substitution of the
equations~\eqref{RelationSigmaPlusWithTheRadialReductionConstraint}
and~\eqref{RescaledVarsInTermsOfReductionConstraints}, into
equation ~\eqref{AsymptsysEFEfirstOrder} and some rearranging reveals,
\begin{flalign}
& (2\mathring{\nabla}_{s} -
H_{LL}{}\mathring{\nabla}_{q})\mathring{\nabla}_{q}H_{ab} \simeq 2
\mathbb{T}_{ab} {} - {}
\gamma_{2}{} \mathbb{C}_{R}{}_{ab} {} - {}
\mathring{\nabla}_{q}H_{aL}\mathring{\nabla}_{q}H_{bL} \nonumber\\
& \qquad\qquad  + 2L_{(a}
\mathring{\nabla}_{|q|}H_{b)}{}^{c}\mathring{\nabla}_{q}H_{Lc} -
\tfrac{1}{2} L_{a} L_{b}
\mathring{\nabla}_{q}H^{cd}\mathring{\nabla}_{q}H_{cd}, \nonumber \\ &
\mathring{\nabla}_qH_{ab} + \tfrac{1}{2}\Sigma^{-}_{ab} \simeq 0,\nonumber\\
& \mathring{\nabla}_q\mathbb{C}_{Rab} \simeq
\gamma_2\mathbb{C}_{Rab}, \nonumber \\ & \mathring{\nabla}_q\mathbb{C}_{S_ A
    ab} \simeq \gamma_2\mathbb{C}_{S_ A ab}.
\end{flalign}
Setting~$\gamma_2 \sim R^{-1}$
and~$\mathcal{T}_{ab}={}^{\mathscr{I}}\mathcal{T}_{ab}$ as given in
equation~\eqref{ConstraintAdditionTotal} and, following the philosophy
of the asymptotic system by formally replacing the~$\simeq$ by~$=$ one
obtains
\begin{flalign}
& (2 \mathring{\nabla}_{s} - H_{LL}{}
\mathring{\nabla}_{q})\mathring{\nabla}_{q}H_{ab} = T_{ab},
\nonumber \\ &
\mathring{\nabla}_qH_{ab} + \tfrac{1}{2}\Sigma^{-}_{ab} =
0, \nonumber \\ &
\mathring{\nabla}_q\mathbb{C}_{Rab} =
0, \nonumber  \\ &
\mathring{\nabla}_q\mathbb{C}_{S_ A ab}
=0, \label{AsymSysFirstOrderInAppropriateVariables}
\end{flalign}
where~$T_{ab}$ is the same tensor as that of
equation~\eqref{eq:AsymptSystemGasHil}.  Finally,
from the last two equations in
~\eqref{AsymSysFirstOrderInAppropriateVariables} one concludes
that
\begin{align}
 \mathbb{C}_{Rab}{} =\mathbb{C}^\star_{Rab} , \quad
 \mathbb{C}_{S_ A}{}_{ab} = \mathbb{C}^\star_{S_ A}{}_{ab},
\end{align}
where~$\mathbb{C}^\star_{R} = \mathbb{C}_{R}|_{q=q_\star},
\mathbb{C}^\star_{S_ A}=\mathbb{C}_{S_ A}|_{q=q_\star}$. Thus,
assuming that the latter quantities are uniformly bounded, it follows
from the analysis of subsection~\ref{sec:AnaAsympSys} and
expressions~\eqref{RescaledVarsInTermsOfReductionConstraints} and
\eqref{RescaledVariablesEFE} that 
\begin{align}
  \sigma^{-}_{\mathcal{U}} & \sim O( R^{-1-\omega}), &
  \sigma^{-}_{\mathcal{V}} & \sim  O( R^{-1}),
  \nonumber \\ \sigma^{+}_{\mathcal{U}} & \sim  O( R^{-2}) , &
  \sigma^{+}_{\mathcal{V}} & \sim  O(R^{-2}), \nonumber \\
    \phi_{S_ A}{}_{\mathcal{U}} & \sim  O(R^{-2}), &
    \phi_{S_ A}{}_{\mathcal{V}} & \sim  O(R^{-2}),
       \nonumber \\
      h_{\mathcal{U}} & \sim  O( R^{-1}) & h_{\mathcal{V}} & \sim  O(R^{-1})
\nonumber \\
 \sigma^{-}_{\underline{L}\underline{L}} & \sim  O( R^{-1}\ln R), &
 \sigma^{+}_{\underline{L}\underline{L}} & \sim  O(R^{-2}\ln R), &
 \nonumber \\ \phi_{S_ A}{}_{\underline{L}\underline{L}} & \sim  O(R^{-2}\ln R), &
 h_{\underline{L}\underline{L}} & \sim  O(R^{-1}\ln R),
 \end{align}
where~$\mathcal{U} \in \{LL, LS_ A, \varnothing \}$, $\mathcal{V}
\in\{ \ L\underline{L}, \underline{L}S_ A, \times, + \}$.

For completeness notice that the generalized harmonic gauge
condition~\eqref{GHGconstraints} implies, to leading order, the
following constraint equation
\begin{align}\label{GHGconstraintPerturbation}
m^{ab}{\Gamma}_{cab} + F_a \doteq 0.
\end{align}
Assuming as before that~$F_a \sim R^{-3}$ a direct calculation shows
that the asymptotic form of these equations read
\begin{align}
\Sigma^{-}_{\mathcal{U}}=0.
\end{align}
Naturally, this implies that when the constraints are not violated we
have~$\sigma^-_{\mathcal{U}}=0$. Nonetheless, recall that, in the
analysis of subsection~\ref{sec:AnaAsympSys} does not rely on the
satisfaction of the asymptotic GHG constraints. In other words, the
analysis of subsection~\ref{sec:AnaAsympSys} shows that even if small
violations of the GHG constraints are present one
has~$\sigma^{-}_{\mathcal{U}}( \gamma) \sim O( R^{-1-\omega})$ so that
the coordinate light-speed condition can be satisfied close
to~$\mathscr{I}$.

%%%%%%%%%%%%%%%%%%%%%%%%%%%%%%%%%%%%%%%%%%%%%%%%%%%%%%%%%%%%%%%%%%%%%%%%%%%%%%%%%%%%%%%
\section{The Trautman-Bondi mass}
%%%%%%%%%%%%%%%%%%%%%%%%%%%%%%%%%%%%%%%%%%%%%%%%%%%%%%%%%%%%%%%%%%%%%%%%%%%%%%%%%%%%%%%

In this section we discuss the implications of the present analysis
for the definition of the Trautman-Bondi mass. The main issue to be
analyzed here is if the~$H_{\underline{L}\underline{L}}$ component
enters into the expression defining the Bondi mass and whether or not
this affects its boundedness.

Let~$\mathcal{H}$ be an hyperboloidal slice in~$(\mathcal{M},g_{ab})$
and let~$\mathcal{S}^2\subset \mathcal{H}$ denote a surface of
constant~$R$ in $\mathcal{H}$. Recall the definitions introduced in
section~\ref{sec:coordlight-speed} for the~$2+1+1$ split and define
\begin{align}
k \equiv q^{ab}\nabla_{a}R_{b}.
\end{align}
where $q^{ab}= g^{ac}g^{bd}q_{cd}$.  Additionally,
let~$\mathring{q}_{ab} \equiv
\mathring{\gamma}_{ab}-\mathring{R}_a\mathring{R}_b$
and~$\mathring{k}\equiv \mathring{q}^{ab}\nabla_{a}\mathring{R}_{b}$,
where $\mathring{q}^{ab}= m^{ac}m^{bd}\mathring{q}_{cd}$,
with~$\mathring{\gamma}_{ab}$ and~$\mathring{R}_a$ as defined in
section~\ref{sec:AsympSysFirstOrderGHG}. Using the above definitions,
the Trautman-Bondi mass can be expressed as
\begin{align}
M=-\frac{1}{8\pi}\int_{\mathcal{S}_{\infty}}(k-\mathring{k})\mbox{d}\mathcal{S}^2
\end{align}
where~$\mbox{d}\mathcal{S}^2$ denotes the area element
in~$(\mathcal{S}^2,q_{ab})$ and~$\mathcal{S}_{\infty}$ represents a
cut of null-infinity~\cite{Poi04}. Consistent with the notation of
subsection~\eqref{sec:coordlight-speed}, some introducing arbitrary
coordinates~$\theta^A$ on~$\mathcal{S}^2$, one
has~$\mbox{d}\mathcal{S}^2=\sqrt{\det[q_{AB}]}\mbox{d}^2\theta$. A
direct computation using equation~\eqref{qABexpansion} gives
\begin{align}\label{ExpansionAreaElement}
\mbox{d}S \simeq R^2 \sqrt{\det[\sigma_{AB}]}\mbox{d}^2\theta
\end{align}
where~$\sigma_{AB}$ denotes the standard metric on~$\mathbb{S}^2$ in
the~$\theta^A$ coordinates. To have a more compact notation we denote
the area element of~$\mathbb{S}^2$ as~$\mbox{d}\mathbb{S}^2$. Observe
that, proceeding in analogous way as for the vector $N^a$ in
section~\ref{sec:AsympSysFirstOrderGHG}, one has the following
expansions
\begin{align}
  S_a \doteq \mathring{R}_a+\check{R}_a, \qquad S^a \doteq
  \mathring{R}^a + (\check{R}^{a}-h^{ab}\mathring{R}_b)
\end{align}
where,
\begin{align}
\check{R}_a \equiv (-\mathring{N}_a\mathring{N}^b\mathring{R}^c +
\tfrac{1}{2}\mathring{R}_a\mathring{R}^b\mathring{R}^c) h_{bc}.
\end{align}
Then, a direct computation shows that
\begin{align}
  k-\mathring{k} \simeq \mathcal{K}+\tfrac{1}{2}C^{\underline{L}}
  -\tfrac{1}{2}\delta^{ A B}\mathring{\nabla}_a
  \Big(\frac{1}{R}H_{\underline{L}S_{ A}}\omega_ B{}^a\Big),
  \label{BondiIntegrand}
\end{align}
where
\begin{flalign}\label{eq:EffectiveIntegrandBondi}
  \mathcal{K} =
  -\frac{H_{LL}}{4R^2}-\frac{H_\varnothing}{4R^2}
  +\frac{\partial_sH^{\varnothing}}{4R^2} +
  \frac{1}{8R^2}(2\partial_s - H_{LL}\partial_q)H_{\underline{LL}}  .
\end{flalign}
and~$C^{\underline{L}} \equiv \underline{\hat{L}}_aC^a$ where $C^a$ encodes
the GHG constraints~\eqref{GHGconstraints}.  Observe that, the last term
in~\eqref{BondiIntegrand} denotes the sum of the divergence of the
vector fields~$\frac{1}{R}H_{\underline{L}S_{ A}}\omega_{ B}{}^a$
which can be regarded as vectors on~$\mathcal{S}^2$. Consequently, the
last term in~\eqref{BondiIntegrand} drops out after integration
on~$\mathcal{S}^2$.

Recall that in the calculation of the asymptotic constraint
conditions~\eqref{eq:AsymptHarmonicConditionComps} were obtained from
the coefficient of the leading order term~$R^{-1}$. Nonetheless,
computed to second order~$C^{\underline{L}}$ reads
\begin{flalign}\label{CLtosecondOrder}
   & C^{\underline{L}} \simeq \frac{\partial_{q}H^{\varnothing}{}}{R}  +
  \frac{1}{R^2}\bigg[ H_{L\underline{L}}{} - \tfrac{1}{2}
  H_{\underline{L} \underline{L}} {} + \tfrac{1}{2}H^{\varnothing}{} +
  {} H_{\underline{L} S_ A}\mathring{\nabla}^a\omega^ A{}_a \nonumber
  \\ & + \delta^{ A B} \omega_{ A}{}^{a}
  \mathring{\nabla}_{a}H_{\underline{L} S_{ B}} +
  \tfrac{1}{2}H_{L\underline{L}}\partial_{q}H^{\varnothing} +
  \tfrac{1}{2} \partial_{s}(H^{\varnothing} -H_{\underline{L}
    \underline{L}}) \nonumber \\ & - {} \partial_q
  \Big((H^\times)^2{} +
  (H^+)^2 + \tfrac{1}{4}(H^{\varnothing})^2 - \tfrac{1}{4}
  H_{LL}H_{\underline{LL}} \Big) \bigg]
\end{flalign}
where it was assumed as usual that~$F_a\sim R^{-3}$ so that the gauge
source functions do not appear in the latter expansion. To simplify
the subsequent discussion, from this point on-wards it will be assumed
that the GHG constraints are satisfied to all orders so
that~$C^a=0$. Taking into account the latter observation one has
\begin{align}
M= -\frac{1}{8\pi}\int_{\mathcal{S}_{\infty}}\mathcal{K}\mbox{d}S.
\end{align}
To verify that the derived expression for~$\mathcal{K}$ is correct
observe that the mass loss formula can be recovered as follows.  Using
that $\partial_T=-\partial_q$ one obtains
\begin{align}
\partial_T M =
\frac{1}{8\pi}\int_{\mathcal{S}_{\infty}}\partial_q\mathcal{K}\mbox{d}S.
\end{align}
Then, a direct computation using~$C^a=0$ to replace~$\partial_q
H_{LL}$ and $\partial_q H^{\varnothing}$ and using
equation~\eqref{ExpansionAreaElement}, renders
\begin{align}
\partial_T M = \frac{1}{64\pi}\int_{\mathbb{S}^2} (2 \partial_{s} -
H_{LL}{} \partial_{q})\partial_{q}H_{\underline{L}\underline{L}}
\mbox{d}\mathbb{S}^2.
\end{align}
Using the asymptotic equation~\eqref{eq:thebad}, one recovers the mass
loss formula
\begin{align}
 \partial_T M = -\frac{1}{16\pi}\int_{\mathbb{S}^2} \bigg(
 (\partial_qH^{+})^2+ (\partial_qH^{\times})^2
 \bigg)\mbox{d}\mathbb{S}^2.
\end{align}
As a side remark it is observed that the Hamiltonian and momentum
constraints an be written in term of the GHG constraints and the
asymptotic equations for~$H_{\underline{L}\underline{L}}$
and~$H_{\underline{L}S_A}$.
 
Having verified that one can recover the mass loss formula, the main
observation to be made is that~$\mathcal{K}$
contains~$H_{\underline{L}\underline{L}}$ and, as discussed in
subsection~\ref{sec:AnaAsympSys}, $H_{\underline{L}\underline{L}} \sim\ln
R$. Nevertheless, the Trautman-Bondi mass is well defined. To see why
this is the case, observe that, since~$C^a=0$ and
hence~$\partial_qH_{LL}=0$, equation~\eqref{eq:thebad} can be written
as
\begin{flalign}
& (2\partial_s-H_{LL}\partial_q)H_{\underline{L}\underline{L}}=
-4\int_{q_\star}^{q} (\partial_qH^{+})^2 + (\partial_qH^{\times})^2
\mbox{d}\bar{q}.
\end{flalign}
Substituting the latter expression into
equation~\eqref{eq:EffectiveIntegrandBondi}, one finds
that~$\mathcal{K}$ can be rewritten as
\begin{multline}
\mathcal{K}= \frac{1}{4R^2}\bigg( -H_{LL}-H^{\varnothing}
+ \partial_sH^{\varnothing} \bigg)  \\
-\frac{1}{2R^2}\int_{q_\star}^{q} (\partial_qH^{+})^2
+(\partial_qH^{\times})^2 \mbox{d}\bar{q}.
\end{multline}
Thus, we observe that, despite that at first instance one would
conclude that~$M$ is diverging as it
contains~$H_{\underline{L}\underline{L}}$, by virtue of the Einstein
field equations, in this case, in the form of the asymptotic equation
for~$H_{\underline{L}\underline{L}}$, the Trautman-Bondi mass is
well defined.

Our final remark is that one can formally recover these results for
the case in which there are small violations to the
constraints~$C^a\neq 0$ simply by redefining the Trautman-Bondi mass
as
\begin{align}
  M=-\frac{1}{8\pi}\int_{\mathcal{S}_{\infty}}(k-\mathring{k})
  - \tfrac{1}{2}C^{\underline{L}} \mbox{d}\mathcal{S}^2
\end{align}
and exploiting the results of subsection~\ref{sec:AnaAsympSys}, to
use~$\partial_qH_{\mathcal{U}} \sim O(R^{-\omega})$ instead
of~$\partial_qH_{\mathcal{U}}=0$, in each of the previous
computations.

%%%%%%%%%%%%%%%%%%%%%%%%%%%%%%%%%%%%%%%%%%%%%%%%%%%%%%%%%%%%%%%%%%%%%%%%%%%%%%%%%%%%%%%
\section{Conclusions}
%%%%%%%%%%%%%%%%%%%%%%%%%%%%%%%%%%%%%%%%%%%%%%%%%%%%%%%%%%%%%%%%%%%%%%%%%%%%%%%%%%%%%%%

The dual foliation
formalism~\cite{HilRic13,Hil15,HilHarBug16,HilRui16,SchHilBug17} is an
approach to GR in which the tensor basis and choice of coordinates are
left uncoupled. In~\cite{HilHarBug16} a proposal was given to use the
formalism to help in the numerical treatment of future null-infinity
via a suitably posed hyperboloidal initial value
problem. Nevertheless, in order for this proposal to work, one of the
requirements, found in~\cite{HilHarBug16}, is that certain derivatives
of the coordinate light-speed have enough decay. The latter condition
is called the coordinate light-speed condition. Here we have studied
 whether or not one can expect the coordinate light-speed condition
to be satisfied. This was done with the use of asymptotic expansions
originally introduced in~\cite{Hor87, Hor97} and employed
in~\cite{LinRod03} to define the weak null condition. We have shown
that the coordinate light-speed condition is related to the asymptotic
harmonic constraints. As discussed in the main text, if the harmonic
constraint condition is satisfied then the coordinate light-speed is
trivially fulfilled. Nevertheless, numerical errors are inherent in
free evolution schemes as one expects, albeit small, violations to the
constraint equations. Moreover, constraint violations are expected to
grow during the numerical evolution. Consequently we must analyze the
system without assuming that the constraints are satisfied. It turns
out that one cannot expect to satisfy the coordinate light-speed
condition without modifying the field equations in such a way that one
damps away constraint violations. Therefore we proposed a constraint
addition such that resulting asymptotic system implies constraint
damping in outgoing null directions. In other words we have shown that
by adding specific multiples of the constraints, the asymptotic system
predicts that the the coordinate light-speed condition will be
satisfied. This paves the way for the explicit numerical treatment of
future null-infinity. Although the constraint addition proposed was
tailored for the purposes of the application in mind, it could be
easily generalized and modified. In the original discussion of the
weak null condition for the Einstein field equations
in~\cite{LinRod03} one can classify the components of the metric
perturbation~$h_{ab}$ into ``good'' components and ``bad''
components. The good components are those whose equations satisfy the
classical null condition of~\cite{Kla80a, Kla86, Chr86} while the
``bad'' component satisfies an equation that fails to satisfy the null
condition. A slower fall-off is hence expected this component. In our
analysis we found that the price to pay to force damping of constraint
violations close to~$\mathscr{I}^+$, and subsequently fulfillment of
the light-speed condition, is to add another layer to this
structure. The components of~$h_{ab}$ are now be classified across
three categories, which we call ``the good, the bad and the
ugly''. The equations for the components lying in the new ``ugly''
category are precisely those associated with the constraints.

Having established fall-off rates within the asymptotic system, we
turned to the Trautman-Bondi mass, and recovered the mass loss
formula. Although at first glance the Trautman-Bondi mass formula
contains terms that could potentially blow up at~$\mathscr{I}^+$, by
virtue of the Einstein field equations it turns out to be well
defined. Finally, in concordance with the outlook of the rest of the 
work, we discussed how to modify the definition of the Trautman-Bondi
with constraints in such a way that one can reproduce the above
remarks, even when small constraints violations are present.

%%%%%%%%%%%%%%%%%%%%%%%%%%%%%%%%%%%%%%%%%%%%%%%%%%%%%%%%%%%%%%%%%%%%%%%%%%%%%%%%%%%%%%%
\acknowledgments
%%%%%%%%%%%%%%%%%%%%%%%%%%%%%%%%%%%%%%%%%%%%%%%%%%%%%%%%%%%%%%%%%%%%%%%%%%%%%%%%%%%%%%%

We are grateful to Laura Bernard, Naresh Dadhich, Adri\'an del R\'io
Vega, Guillaume Faye, Chris Moore, Shalabh, Juan Valiente Kroon and
Alex Va\~n\'o-Vi\~nuales for helpful discussions. DH also gratefully
acknowledges support offered by IUCAA, Pune, where part of this work
was completed. The work was partially supported by the FCT (Portugal)
IF Program~IF/00577/2015.

%%%%%%%%%%%%%%%%%%%%%%%%%%%%%%%%%%%%%%%%%%%%%%%%%%%%%%%%%%%%%%%%%%%%%%%%%%%%%%%%%%%%%%%
\bibliographystyle{unsrt}
\bibliography{As_Hyp_DF.bbl}{}

\begin{thebibliography}{46}
\expandafter\ifx\csname natexlab\endcsname\relax\def\natexlab#1{#1}\fi
\expandafter\ifx\csname bibnamefont\endcsname\relax
  \def\bibnamefont#1{#1}\fi
\expandafter\ifx\csname bibfnamefont\endcsname\relax
  \def\bibfnamefont#1{#1}\fi
\expandafter\ifx\csname citenamefont\endcsname\relax
  \def\citenamefont#1{#1}\fi
\expandafter\ifx\csname url\endcsname\relax
  \def\url#1{\texttt{#1}}\fi
\expandafter\ifx\csname urlprefix\endcsname\relax\def\urlprefix{URL }\fi
\providecommand{\bibinfo}[2]{#2}
\providecommand{\eprint}[2][]{\url{#2}}

\bibitem[{\citenamefont{Penrose}(1963)}]{Pen63}
\bibinfo{author}{\bibfnamefont{R.}~\bibnamefont{Penrose}},
  \bibinfo{journal}{Phys. Rev. Lett.} \textbf{\bibinfo{volume}{10}},
  \bibinfo{pages}{66} (\bibinfo{year}{1963}).

\bibitem[{\citenamefont{Penrose}(1964)}]{Pen64}
\bibinfo{author}{\bibfnamefont{R.}~\bibnamefont{Penrose}}, in
  \emph{\bibinfo{booktitle}{Relativity, Groups, and Topology (Les Houches,
  France, 1964)}}, edited by
  \bibinfo{editor}{\bibfnamefont{C.}~\bibnamefont{DeWitt}} \bibnamefont{and}
  \bibinfo{editor}{\bibfnamefont{B.}~\bibnamefont{DeWitt}}
  (\bibinfo{publisher}{Gordon and Breach}, \bibinfo{address}{New York},
  \bibinfo{year}{1964}), pp. \bibinfo{pages}{565--584}.

\bibitem[{\citenamefont{Husa}(2002)}]{Hus02}
\bibinfo{author}{\bibfnamefont{S.}~\bibnamefont{Husa}}, in
  \emph{\bibinfo{booktitle}{The Conformal Structure of Spacetimes: Geometry,
  Analysis, Numerics}}, edited by
  \bibinfo{editor}{\bibfnamefont{J.}~\bibnamefont{Frauendiener}}
  \bibnamefont{and} \bibinfo{editor}{\bibfnamefont{H.}~\bibnamefont{Friedrich}}
  (\bibinfo{year}{2002}), vol. \bibinfo{volume}{604}, chap.
  \bibinfo{chapter}{Problems and {S}uccesses in the {N}umerical {A}pproach to
  the {C}onformal {F}ield {E}quations}.

\bibitem[{\citenamefont{Stewart}(1991)}]{Ste91}
\bibinfo{author}{\bibfnamefont{J.}~\bibnamefont{Stewart}},
  \emph{\bibinfo{title}{Advanced general relativity}}
  (\bibinfo{publisher}{Cambridge University Press}, \bibinfo{year}{1991}).

\bibitem[{\citenamefont{Frauendiener}(2004)}]{Fra04}
\bibinfo{author}{\bibfnamefont{J.}~\bibnamefont{Frauendiener}},
  \bibinfo{journal}{Living Rev. Relativity} \textbf{\bibinfo{volume}{7}}
  (\bibinfo{year}{2004}).

\bibitem[{\citenamefont{{Ashtekar}}(2014)}]{Ash14}
\bibinfo{author}{\bibfnamefont{A.}~\bibnamefont{{Ashtekar}}},
  \bibinfo{journal}{ArXiv e-prints}  (\bibinfo{year}{2014}),
  \eprint{1409.1800}.

\bibitem[{\citenamefont{Valiente-Kroon}(2016)}]{Val16}
\bibinfo{author}{\bibfnamefont{J.-A.} \bibnamefont{Valiente-Kroon}},
  \emph{\bibinfo{title}{Conformal Methods in General Relativity}}
  (\bibinfo{publisher}{Cambridge University Press},
  \bibinfo{address}{Cambridge}, \bibinfo{year}{2016}).

\bibitem[{\citenamefont{Trautman}(1958)}]{Tra58a}
\bibinfo{author}{\bibfnamefont{A.}~\bibnamefont{Trautman}},
  \bibinfo{journal}{Bulletin of the Polish Academy of Sciences}
  \textbf{\bibinfo{volume}{VI}}, \bibinfo{pages}{407} (\bibinfo{year}{1958}).

\bibitem[{\citenamefont{Bondi et~al.}(1962)\citenamefont{Bondi, van~der Burg,
  and Metzner}}]{BonBurMet62}
\bibinfo{author}{\bibfnamefont{H.}~\bibnamefont{Bondi}},
  \bibinfo{author}{\bibfnamefont{M.~G.~J.} \bibnamefont{van~der Burg}},
  \bibnamefont{and} \bibinfo{author}{\bibfnamefont{A.~W.~K.}
  \bibnamefont{Metzner}}, \bibinfo{journal}{Proc. Roy. Soc. A}
  \textbf{\bibinfo{volume}{269}}, \bibinfo{pages}{21} (\bibinfo{year}{1962}).

\bibitem[{\citenamefont{Sachs}(1962)}]{Sac62a}
\bibinfo{author}{\bibfnamefont{R.}~\bibnamefont{Sachs}},
  \bibinfo{journal}{Proc. Roy. Soc. London} \textbf{\bibinfo{volume}{A270}},
  \bibinfo{pages}{103} (\bibinfo{year}{1962}).

\bibitem[{\citenamefont{Abbott et~al.}(2016)}]{AbbAbbAbb16}
\bibinfo{author}{\bibfnamefont{B.~P.} \bibnamefont{Abbott}}
  \bibnamefont{et~al.}, \bibinfo{journal}{Phys. Rev. Lett.}
  \textbf{\bibinfo{volume}{116}}, \bibinfo{pages}{061102}
  (\bibinfo{year}{2016}), \eprint{1602.03837}.

\bibitem[{\citenamefont{Winicour}(2012)}]{Win12}
\bibinfo{author}{\bibfnamefont{J.}~\bibnamefont{Winicour}},
  \bibinfo{journal}{Living Rev. Relativity} \textbf{\bibinfo{volume}{15}},
  \bibinfo{pages}{2} (\bibinfo{year}{2012}), \bibinfo{note}{[Online article]},
  \urlprefix\url{http://www.livingreviews.org/lrr-2012-2}.

\bibitem[{\citenamefont{Friedrich}(1981)}]{Fri81}
\bibinfo{author}{\bibfnamefont{H.}~\bibnamefont{Friedrich}},
  \bibinfo{journal}{Proc. Roy. Soc. London} \textbf{\bibinfo{volume}{A 375}},
  \bibinfo{pages}{169} (\bibinfo{year}{1981}).

\bibitem[{\citenamefont{Doulis and Frauendiener}(2016)}]{DouFra16}
\bibinfo{author}{\bibfnamefont{G.}~\bibnamefont{Doulis}} \bibnamefont{and}
  \bibinfo{author}{\bibfnamefont{J.}~\bibnamefont{Frauendiener}}
  (\bibinfo{year}{2016}), \eprint{1609.03584}.

\bibitem[{\citenamefont{Zenginoglu}(2008)}]{Zen08}
\bibinfo{author}{\bibfnamefont{A.}~\bibnamefont{Zenginoglu}},
  \bibinfo{journal}{Class. Quant. Grav.} \textbf{\bibinfo{volume}{25}},
  \bibinfo{pages}{195025} (\bibinfo{year}{2008}), \eprint{0808.0810}.

\bibitem[{\citenamefont{Va{\~n}{\'o}-Vi{\~n}uales
  et~al.}(2015)\citenamefont{Va{\~n}{\'o}-Vi{\~n}uales, Husa, and
  Hilditch}}]{VanHusHil14}
\bibinfo{author}{\bibfnamefont{A.}~\bibnamefont{Va{\~n}{\'o}-Vi{\~n}uales}},
  \bibinfo{author}{\bibfnamefont{S.}~\bibnamefont{Husa}}, \bibnamefont{and}
  \bibinfo{author}{\bibfnamefont{D.}~\bibnamefont{Hilditch}},
  \bibinfo{journal}{Class. Quant. Grav.} \textbf{\bibinfo{volume}{32}},
  \bibinfo{pages}{175010} (\bibinfo{year}{2015}), \eprint{1412.3827}.

\bibitem[{\citenamefont{Va{\~n}{\'o}-Vi{\~n}uales and Husa}(2015)}]{VanHus14}
\bibinfo{author}{\bibfnamefont{A.}~\bibnamefont{Va{\~n}{\'o}-Vi{\~n}uales}}
  \bibnamefont{and} \bibinfo{author}{\bibfnamefont{S.}~\bibnamefont{Husa}},
  \bibinfo{journal}{J. Phys. Conf. Ser.} \textbf{\bibinfo{volume}{600}},
  \bibinfo{pages}{012061} (\bibinfo{year}{2015}), \eprint{1412.4801}.

\bibitem[{\citenamefont{Va{\~n}{\'o}-Vi{\~n}uale}(2015)}]{Van15}
\bibinfo{author}{\bibfnamefont{A.}~\bibnamefont{Va{\~n}{\'o}-Vi{\~n}uale}},
  Ph.D. thesis, \bibinfo{school}{U. Iles Balears, Palma}
  (\bibinfo{year}{2015}), \eprint{1512.00776},
  \urlprefix\url{http://inspirehep.net/record/1407828/files/arXiv:1512.00776.pdf}.

\bibitem[{\citenamefont{Moncrief and Rinne}(2009)}]{MonRin08}
\bibinfo{author}{\bibfnamefont{V.}~\bibnamefont{Moncrief}} \bibnamefont{and}
  \bibinfo{author}{\bibfnamefont{O.}~\bibnamefont{Rinne}},
  \bibinfo{journal}{Class.Quant.Grav.} \textbf{\bibinfo{volume}{26}},
  \bibinfo{pages}{125010} (\bibinfo{year}{2009}), \eprint{0811.4109}.

\bibitem[{\citenamefont{Bardeen et~al.}(2011)\citenamefont{Bardeen, Sarbach,
  and Buchman}}]{BarSarBuc11}
\bibinfo{author}{\bibfnamefont{J.~M.} \bibnamefont{Bardeen}},
  \bibinfo{author}{\bibfnamefont{O.}~\bibnamefont{Sarbach}}, \bibnamefont{and}
  \bibinfo{author}{\bibfnamefont{L.~T.} \bibnamefont{Buchman}},
  \bibinfo{journal}{Phys. Rev.} \textbf{\bibinfo{volume}{D83}},
  \bibinfo{pages}{104045} (\bibinfo{year}{2011}), \eprint{1101.5479}.

\bibitem[{\citenamefont{Calabrese et~al.}(2006)\citenamefont{Calabrese,
  Gundlach, and Hilditch}}]{CalGunHil05}
\bibinfo{author}{\bibfnamefont{G.}~\bibnamefont{Calabrese}},
  \bibinfo{author}{\bibfnamefont{C.}~\bibnamefont{Gundlach}}, \bibnamefont{and}
  \bibinfo{author}{\bibfnamefont{D.}~\bibnamefont{Hilditch}},
  \bibinfo{journal}{Class.Quant.Grav.} \textbf{\bibinfo{volume}{23}},
  \bibinfo{pages}{4829} (\bibinfo{year}{2006}), \eprint{gr-qc/0512149}.

\bibitem[{\citenamefont{Hilditch et~al.}(2018)\citenamefont{Hilditch, Harms,
  Bugner, R{\"u}ter, and Br{\"u}gmann}}]{HilHarBug16}
\bibinfo{author}{\bibfnamefont{D.}~\bibnamefont{Hilditch}},
  \bibinfo{author}{\bibfnamefont{E.}~\bibnamefont{Harms}},
  \bibinfo{author}{\bibfnamefont{M.}~\bibnamefont{Bugner}},
  \bibinfo{author}{\bibfnamefont{H.}~\bibnamefont{R{\"u}ter}},
  \bibnamefont{and}
  \bibinfo{author}{\bibfnamefont{B.}~\bibnamefont{Br{\"u}gmann}},
  \bibinfo{journal}{Class. Quant. Grav.} \textbf{\bibinfo{volume}{35}},
  \bibinfo{pages}{055003} (\bibinfo{year}{2018}), \eprint{1609.08949}.

\bibitem[{\citenamefont{Hilditch}(2015)}]{Hil15}
\bibinfo{author}{\bibfnamefont{D.}~\bibnamefont{Hilditch}}
  (\bibinfo{year}{2015}), \eprint{1509.02071}.

\bibitem[{\citenamefont{Lindblad and Rodnianski}(2003)}]{LinRod03}
\bibinfo{author}{\bibfnamefont{H.}~\bibnamefont{Lindblad}} \bibnamefont{and}
  \bibinfo{author}{\bibfnamefont{I.}~\bibnamefont{Rodnianski}},
  \bibinfo{journal}{Comptes Rendus Mathematique}
  \textbf{\bibinfo{volume}{336}}, \bibinfo{pages}{901 } (\bibinfo{year}{2003}),
  ISSN \bibinfo{issn}{1631-073X},
  \urlprefix\url{http://www.sciencedirect.com/science/article/pii/S1631073X03002310}.

\bibitem[{\citenamefont{{Keir}}(2018)}]{Kei17}
\bibinfo{author}{\bibfnamefont{J.}~\bibnamefont{{Keir}}},
  \bibinfo{journal}{ArXiv e-prints}  (\bibinfo{year}{2018}),
  \eprint{1808.09982}.

\bibitem[{\citenamefont{Brodbeck et~al.}(1999)\citenamefont{Brodbeck,
  Frittelli, H{\"u}bner, and Reula}}]{BroFriHub98}
\bibinfo{author}{\bibfnamefont{O.}~\bibnamefont{Brodbeck}},
  \bibinfo{author}{\bibfnamefont{S.}~\bibnamefont{Frittelli}},
  \bibinfo{author}{\bibfnamefont{P.}~\bibnamefont{H{\"u}bner}},
  \bibnamefont{and} \bibinfo{author}{\bibfnamefont{O.~A.} \bibnamefont{Reula}},
  \bibinfo{journal}{J. Math. Phys.} \textbf{\bibinfo{volume}{40}},
  \bibinfo{pages}{909} (\bibinfo{year}{1999}), \eprint{gr-qc/9809023}.

\bibitem[{\citenamefont{Gundlach et~al.}(2005)\citenamefont{Gundlach,
  Martin-Garcia, Calabrese, and Hinder}}]{GunGarCal05}
\bibinfo{author}{\bibfnamefont{C.}~\bibnamefont{Gundlach}},
  \bibinfo{author}{\bibfnamefont{J.~M.} \bibnamefont{Martin-Garcia}},
  \bibinfo{author}{\bibfnamefont{G.}~\bibnamefont{Calabrese}},
  \bibnamefont{and} \bibinfo{author}{\bibfnamefont{I.}~\bibnamefont{Hinder}},
  \bibinfo{journal}{Class. Quantum Grav.} \textbf{\bibinfo{volume}{22}},
  \bibinfo{pages}{3767} (\bibinfo{year}{2005}), \eprint{gr-qc/0504114}.

\bibitem[{\citenamefont{Lindblom et~al.}(2006)\citenamefont{Lindblom, Scheel,
  Kidder, Owen, and Rinne}}]{LinSchKid05}
\bibinfo{author}{\bibfnamefont{L.}~\bibnamefont{Lindblom}},
  \bibinfo{author}{\bibfnamefont{M.~A.} \bibnamefont{Scheel}},
  \bibinfo{author}{\bibfnamefont{L.~E.} \bibnamefont{Kidder}},
  \bibinfo{author}{\bibfnamefont{R.}~\bibnamefont{Owen}}, \bibnamefont{and}
  \bibinfo{author}{\bibfnamefont{O.}~\bibnamefont{Rinne}},
  \bibinfo{journal}{Class. Quant. Grav.} \textbf{\bibinfo{volume}{23}},
  \bibinfo{pages}{S447} (\bibinfo{year}{2006}), \eprint{gr-qc/0512093}.

\bibitem[{\citenamefont{H{\"o}rmander}(1987)}]{Hor87}
\bibinfo{author}{\bibfnamefont{L.}~\bibnamefont{H{\"o}rmander}},
  \emph{\bibinfo{title}{The lifespan of classical solutions of non-linear
  hyperbolic equations}} (\bibinfo{publisher}{Springer Berlin Heidelberg},
  \bibinfo{address}{Berlin, Heidelberg}, \bibinfo{year}{1987}), pp.
  \bibinfo{pages}{214--280}, ISBN \bibinfo{isbn}{978-3-540-47886-7},
  \urlprefix\url{https://doi.org/10.1007/BFb0077745}.

\bibitem[{\citenamefont{H{\"o}rmander}(1997)}]{Hor97}
\bibinfo{author}{\bibfnamefont{L.}~\bibnamefont{H{\"o}rmander}},
  \emph{\bibinfo{title}{Lectures on Nonlinear Hyperbolic Differential
  Equations}}, Math{\'e}matiques et Applications (\bibinfo{publisher}{Springer
  Berlin Heidelberg}, \bibinfo{year}{1997}), ISBN
  \bibinfo{isbn}{9783540629214},
  \urlprefix\url{https://books.google.pt/books?id=qps02wnhmEMC}.

\bibitem[{\citenamefont{Sogge}(1995)}]{Sog95}
\bibinfo{author}{\bibfnamefont{C.}~\bibnamefont{Sogge}},
  \emph{\bibinfo{title}{Lectures on nonlinear wave equations}}, no.
  \bibinfo{number}{Bd. 2} in \bibinfo{series}{Monographs in analysis}
  (\bibinfo{publisher}{International Press}, \bibinfo{year}{1995}).

\bibitem[{\citenamefont{{Lindblad} and {Rodnianski}}(2005)}]{LinRod05}
\bibinfo{author}{\bibfnamefont{H.}~\bibnamefont{{Lindblad}}} \bibnamefont{and}
  \bibinfo{author}{\bibfnamefont{I.}~\bibnamefont{{Rodnianski}}},
  \bibinfo{journal}{Communications in Mathematical Physics}
  \textbf{\bibinfo{volume}{256}}, \bibinfo{pages}{43} (\bibinfo{year}{2005}),
  \eprint{math/0312479}.

\bibitem[{\citenamefont{Lindblad}(2017)}]{Lin17}
\bibinfo{author}{\bibfnamefont{H.}~\bibnamefont{Lindblad}},
  \bibinfo{journal}{Communications in Mathematical Physics}
  \textbf{\bibinfo{volume}{353}}, \bibinfo{pages}{135} (\bibinfo{year}{2017}),
  \urlprefix\url{https://doi.org/10.1007/s00220-017-2876-z}.

\bibitem[{\citenamefont{{Lindblad} and {Rodnianski}}(2004)}]{LinRod04}
\bibinfo{author}{\bibfnamefont{H.}~\bibnamefont{{Lindblad}}} \bibnamefont{and}
  \bibinfo{author}{\bibfnamefont{I.}~\bibnamefont{{Rodnianski}}},
  \bibinfo{journal}{ArXiv Mathematics e-prints}  (\bibinfo{year}{2004}),
  \eprint{math/0411109}.

\bibitem[{\citenamefont{Weyhausen et~al.}(2012)\citenamefont{Weyhausen,
  Bernuzzi, and Hilditch}}]{WeyBerHil11}
\bibinfo{author}{\bibfnamefont{A.}~\bibnamefont{Weyhausen}},
  \bibinfo{author}{\bibfnamefont{S.}~\bibnamefont{Bernuzzi}}, \bibnamefont{and}
  \bibinfo{author}{\bibfnamefont{D.}~\bibnamefont{Hilditch}},
  \bibinfo{journal}{Phys. Rev. D} \textbf{\bibinfo{volume}{85}},
  \bibinfo{pages}{024038} (\bibinfo{year}{2012}), \eprint{1107.5539}.

\bibitem[{Gas()}]{GasHil18_web}
\bibinfo{note}{\url{http://www.tpi.uni-jena.de/~hild/AsHypDF.tgz}}.

\bibitem[{\citenamefont{Mart{\'i}n-Garc{\'i}a}(2017)}]{xAct_web_aastex}
\bibinfo{author}{\bibfnamefont{J.~M.} \bibnamefont{Mart{\'i}n-Garc{\'i}a}},
  \emph{\bibinfo{title}{x{A}ct: tensor computer algebra.}}
  (\bibinfo{year}{2017}), \bibinfo{note}{\url{http://www.xact.es/}}.

\bibitem[{\citenamefont{Taylor}(1996)}]{Tay96}
\bibinfo{author}{\bibfnamefont{M.~E.} \bibnamefont{Taylor}},
  \emph{\bibinfo{title}{Partial differential equations {III}: nonlinear
  equations}} (\bibinfo{publisher}{Springer Verlag}, \bibinfo{year}{1996}).

\bibitem[{\citenamefont{{G{\'o}mez-Lobo} and {Valiente
  Kroon}}(2008)}]{GomVal08}
\bibinfo{author}{\bibfnamefont{A.~G.-P.} \bibnamefont{{G{\'o}mez-Lobo}}}
  \bibnamefont{and} \bibinfo{author}{\bibfnamefont{J.~A.}
  \bibnamefont{{Valiente Kroon}}}, \bibinfo{journal}{Journal of Geometry and
  Physics} \textbf{\bibinfo{volume}{58}}, \bibinfo{pages}{1186}
  (\bibinfo{year}{2008}), \eprint{0712.3373}.

\bibitem[{\citenamefont{Poisson}(2004)}]{Poi04}
\bibinfo{author}{\bibfnamefont{E.}~\bibnamefont{Poisson}},
  \emph{\bibinfo{title}{A Relativist's Toolkit: The Mathematics of Black-Hole
  Mechanics}} (\bibinfo{publisher}{Cambridge University Press},
  \bibinfo{year}{2004}), ISBN \bibinfo{isbn}{0521830915}.

\bibitem[{\citenamefont{Hilditch and Richter}(2016)}]{HilRic13}
\bibinfo{author}{\bibfnamefont{D.}~\bibnamefont{Hilditch}} \bibnamefont{and}
  \bibinfo{author}{\bibfnamefont{R.}~\bibnamefont{Richter}},
  \bibinfo{journal}{Phys. Rev.} \textbf{\bibinfo{volume}{D94}},
  \bibinfo{pages}{044028} (\bibinfo{year}{2016}), \eprint{1303.4783}.

\bibitem[{\citenamefont{Hilditch and Ruiz}(2016)}]{HilRui16}
\bibinfo{author}{\bibfnamefont{D.}~\bibnamefont{Hilditch}} \bibnamefont{and}
  \bibinfo{author}{\bibfnamefont{M.}~\bibnamefont{Ruiz}}
  (\bibinfo{year}{2016}), \eprint{1609.06925}.

\bibitem[{\citenamefont{Schoepe et~al.}(2018)\citenamefont{Schoepe, Hilditch,
  and Bugner}}]{SchHilBug17}
\bibinfo{author}{\bibfnamefont{A.}~\bibnamefont{Schoepe}},
  \bibinfo{author}{\bibfnamefont{D.}~\bibnamefont{Hilditch}}, \bibnamefont{and}
  \bibinfo{author}{\bibfnamefont{M.}~\bibnamefont{Bugner}},
  \bibinfo{journal}{Phys. Rev.} \textbf{\bibinfo{volume}{D97}},
  \bibinfo{pages}{123009} (\bibinfo{year}{2018}), \eprint{1712.09837}.

\bibitem[{\citenamefont{Klainerman}(1980)}]{Kla80a}
\bibinfo{author}{\bibfnamefont{S.}~\bibnamefont{Klainerman}},
  \bibinfo{journal}{Communications on Pure and Applied Mathematics}
  \textbf{\bibinfo{volume}{33}}, \bibinfo{pages}{43} (\bibinfo{year}{1980}).

\bibitem[{\citenamefont{Klainerman}(1986)}]{Kla86}
\bibinfo{author}{\bibfnamefont{S.}~\bibnamefont{Klainerman}}, in
  \emph{\bibinfo{booktitle}{Nonlinear systems of partial differential equations
  in applied mathematics, {P}art 1 ({S}anta {F}e, {N}.{M}., 1984)}}
  (\bibinfo{publisher}{Amer. Math. Soc., Providence, RI},
  \bibinfo{year}{1986}), vol.~\bibinfo{volume}{23} of
  \emph{\bibinfo{series}{Lectures in Appl. Math.}}, pp.
  \bibinfo{pages}{293--326}.

\bibitem[{\citenamefont{Christoudolou}(1986)}]{Chr86}
\bibinfo{author}{\bibfnamefont{D.}~\bibnamefont{Christoudolou}},
  \bibinfo{journal}{Communications in Mathematical Physics}
  \textbf{\bibinfo{volume}{105}}, \bibinfo{pages}{337} (\bibinfo{year}{1986}).

\end{thebibliography}
%%%%%%%%%%%%%%%%%%%%%%%%%%%%%%%%%%%%%%%%%%%%%%%%%%%%%%%%%%%%%%%%%%%%%%%%%%%%%%%%%%%%%%%

%%%%%%%%%%%%%%%%%%%%%%%%%%%%%%%%%%%%%%%%%%%%%%%%%%%%%%%%%%%%%%%%%%%%%%%%%%%%%%%%%%%%%%%
\end{document}